\documentclass[]{aastex631}

\pdfoutput=1

\shorttitle{AASTeX v6.3.1 Sample article}
\shortauthors{Zhang et al.}

\graphicspath{{./}{figures/}}

\begin{document}

\title{Optical Observations of the Nearby Type Ia Supernova 2021hpr}

\author[0000-0003-2536-2641]{Yu Zhang}
\affil{Department of Physics, College of Science, Tibet University, Lhasa 850000, People's Republic of China}
\affil{Key Laboratory of Cosmic Rays (Tibet University), Ministry of Education, Lhasa 850000, People's Republic of China}
\affil{Key Laboratory of Optical Astronomy, National Astronomical Observatories, Chinese Academy of Sciences, Beijing 100101, People's Republic of China}
\affil{School of Astronomy and Space Science, University of Chinese Academy of Sciences, Beijing 101408, People's Republic of China}

\author[0000-0002-8531-5161]{Tianmeng Zhang}
\affil{Key Laboratory of Optical Astronomy, National Astronomical Observatories, Chinese Academy of Sciences, Beijing 100101, People's Republic of China}
\affil{School of Astronomy and Space Science, University of Chinese Academy of Sciences, Beijing 101408, People's Republic of China}

\author{Danzengluobu}
\affiliation{Key Laboratory of Cosmic Rays (Tibet University), Ministry of Education, Lhasa 850000, People's Republic of China}
\affiliation{Department of Physics, College of Science, Tibet University, Lhasa 850000, People's Republic of China}

\author[0000-0001-6489-163X]{Zhitong Li}
\affil{Key Laboratory of Optical Astronomy, National Astronomical Observatories, Chinese Academy of Sciences, Beijing 100101, People's Republic of China}
\affil{School of Astronomy and Space Science, University of Chinese Academy of Sciences, Beijing 101408, People's Republic of China}

\author[0000-0002-4328-538X]{Pinsong Zhao}
\affil{Key Laboratory of Optical Astronomy, National Astronomical Observatories, Chinese Academy of Sciences, Beijing 100101, People's Republic of China}
\affil{School of Astronomy and Space Science, University of Chinese Academy of Sciences, Beijing 101408, People's Republic of China}

\author[0000-0002-6659-1152]{Bingqing Zhang}
\affil{Key Laboratory of Optical Astronomy, National Astronomical Observatories, Chinese Academy of Sciences, Beijing 100101, People's Republic of China}
\affil{School of Astronomy and Space Science, University of Chinese Academy of Sciences, Beijing 101408, People's Republic of China}

\author[0000-0002-4915-4137]{Lin Du}
\affil{Key Laboratory of Optical Astronomy, National Astronomical Observatories, Chinese Academy of Sciences, Beijing 100101, People's Republic of China}
\affil{School of Astronomy and Space Science, University of Chinese Academy of Sciences, Beijing 101408, People's Republic of China}

\author{Yinan Zhu}
\affil{Key Laboratory of Optical Astronomy, National Astronomical Observatories, Chinese Academy of Sciences, Beijing 100101, People's Republic of China}

\author{Hong Wu}
\affil{Key Laboratory of Optical Astronomy, National Astronomical Observatories, Chinese Academy of Sciences, Beijing 100101, People's Republic of China}
\affiliation{School of Astronomy and Space Science, University of Chinese Academy of Sciences, Beijing 101408, People's Republic of China}

\correspondingauthor{Tianmeng Zhang}
\email{zhangtm@nao.cas.cn}
\correspondingauthor{Danzengluobu}
\email{dzlb@Utibet.edu.cn}
\correspondingauthor{Hong Wu}
\email{hwu@bao.ac.cn}

\begin{abstract}

We present the optical photometric and spectroscopic observations of the nearby Type Ia supernova (SN) 2021hpr. 
The observations covered the phase of $-$14.37 to +63.68 days relative to its maximum luminosity in the $B$ band. 
The evolution of multiband light/color curves of SN 2021hpr is similar to that of normal Type Ia supernovae (SNe Ia) with the exception of some phases, especially a plateau phase that appeared in the $V-R$ color curve before peak luminosity, which resembles that of SN 2017cbv. 
The first spectrum we observed at t $\sim -$14.4 days shows a higher velocity for the Si {\sc ii} $\lambda$6355 feature ($\sim$ 21,000 km s$^{-1}$) than that of other normal Velocity (NV) SNe Ia at the same phase.
Based on the Si {\sc ii} $\lambda$6355 velocity of $\sim$ 12,420 km s$^{-1}$ around the maximum light, we deduce that SN 2021hpr is a transitional object between high velocity (HV) and NV SNe Ia. 
Meanwhile, the Si {\sc ii} $\lambda$6355 feature shows a high velocity gradient (HVG) of about 800 km s$^{-1}$ day$^{-1}$ from roughly $-14.37$ to $-4.31$ days relative to the $B$-band maximum, which indicates that SN 2021hpr can also be classified as an HVG SN Ia. 
Despite SN 2021hpr having a higher velocity for the Si {\sc ii} $\lambda$6355 and Ca {\sc ii} near-IR (NIR) triplet features in its spectra, its evolution is similar to that of SN 2011fe. 
Including SN 2021hpr, there have been six supernovae observed in the host galaxy NGC 3147, and the supernovae explosion rate in the last 50 yr is slightly higher for SNe Ia, while lower for SNe Ibc and SNe II it is lower than expected rate from the radio data.
Inspecting the spectra, we find that SN 2021hpr has a metal-rich (12 + log(O/H) $\approx$ 8.648) circumstellar environment, where HV SNe tend to reside.
Based on the decline rate of SN 2021hpr in the $B$ band, we determine the distance modulus of the host galaxy NGC 3147 using the Phillips relation to be 33.46 $\pm$ 0.21 mag, which is close to that found by previous works.

\end{abstract}


\section{Introduction} \label{sec:intro}

Supernovae, among of the most violent astrophysical phenomena in the universe, are the endpoint of the evolution of some stars, meaning that those stars completely collapse.
Type Ia supernovae (SNe Ia) have been used as good distance indicators to measure the accelerating expansion of the universe because of their high luminosities and homogeneous properties \citep{1998AJ....116.1009R, 1999ApJ...517..565P, 2016ApJ...826...56R, 2018ApJ...853..126R}. Presently, SNe Ia are widely used as "standard candles" to measure the distances of external galaxies and are a means to determine the Hubble Constant \citep{1996ApJ...473...88R, 2005A&A...443..781G, 2005ApJ...620L..87W, 2017MNRAS.471.2254Z, 2018ApJ...869...56B, 2019ApJ...882...34F, 2020MNRAS.498.1420W}.
Besides, SNe Ia are also produce most of the iron group elements in the universe and drive the evolution of the element abundances in galaxies \citep{1998ApJ...501..643D, 1998ApJ...503L.155K, 2009ApJ...707.1466K, 2016ApJ...824...82C, 2021MNRAS.506.5951L}. 

Two popular scenarios have been proposed for the progenitor of SNe Ia \citep{2014ARA&A..52..107M}: 
one is the single degenerate (SD) scenario \citep{1973ApJ...186.1007W}, in which a C/O white dwarf (WD) explodes when it reaches the Chandrasekhar mass limit by accreting material from its nondegenerate companion star. 
The other is double degenerate (DD) scenario \citep{1984ApJ...277..355W, 1984ApJS...54..335I}, in which the explosion is triggered by the merging of a double degenerate system via accretion or violent collision. 
Both SD \citep{2011Sci...333..856S, 2012Sci...337..942D, 2013ApJS..207....3S, 2013MNRAS.436..222M, 2015A&A...577A..39L, 2018MNRAS.473..336B} and DD \citep{2011Natur.480..348L, 2015Natur.521..328C, 2015Natur.521..332O} scenarios are supported by multiple observational samples.
However, the explosion mechanism of SNe Ia and details of the progenitor system remain controversial \citep{1969Ap&SS...5..180A, 1991A&A...245..114K, 2002A&A...391.1167R, 2004ApJ...612L..37P, 2007ApJ...668.1103R, 2011A&A...528A.117P, 2012ApJ...747L..10P, 2013MNRAS.429.1156S, 2014ARA&A..52..107M, 2016A&A...592A..57S, 2017suex.book.....B, 2020A&A...635A.169G, 2021ApJ...909L..18S}.

\cite{2005ApJ...623.1011B} suggest that normal SNe Ia can be classified into three classes: FAINT SNe Ia, high velocity gradient (HVG) SNe Ia, and low velocity gradient (LVG) SNe Ia, based on the temporal velocity gradient of Si {\sc ii}, while \cite{2009ApJ...699L.139W} suggest that SNe Ia can be divided into two subclasses, normal velocity (NV) SNe Ia and high velocity (HV) SNe Ia, 
based on the Si {\sc ii} $\lambda$6355 expansion velocity near the $B$-band maximum. 
In addition to having ejecta with higher velocities ($\gtrsim$ 11,800 km s$^{-1}$), HV SNe Ia also show a redder peak $B-V$ color, and a slower $B$-band decline rate than those of NV SNe Ia \citep{2009ApJ...699L.139W, 2011ApJ...729...55F, 2011ApJ...742...89F, 2012ApJ...748..127F, 2014ApJ...797...75M, 2014MNRAS.437..338C}. 
As for the differences between these two subclasses, they could result from a geometric effect of explosion asymmetry and viewing angle \citep{2010ApJ...725L.167M, 2010Natur.466...82M, 2011ApJ...732...30P, 2011MNRAS.413.3075M}. 
A recent study by \cite{2019ApJ...873...84P} has another, highly plausible, explanation for the high velocities and redder colors. They suggest that the differences are likely to be a natural byproduct of the double detonation model. 
Furthermore, the HV SNe Ia tend to reside in the large and massive galaxies, and usually arise in the inner, brighter, and metal-rich areas of their host galaxies, based on the analysis of the circumstellar environment of SNe Ia \citep{2013Sci...340..170W, 2015MNRAS.446..354P, 2020ApJ...895L...5P, 2021ApJ...906...99L}.

The unresolved explosion mechanism and the diversity of SNe Ia relative to the circumstellar environment would directly affect the accuracy of the distance measurement of distant objects in the universe. Therefore, more and earlier photometric and spectroscopic data of SNe Ia are necessary.
Fortunately, different models and mechanisms of SNe Ia progenitors exhibit significant differences when observed at their early stages, which is crucial to constrain the progenitors of the SNe Ia. 
The extra emission at early times could be produced by the interaction between the ejecta and the nondegenerate companion star \citep{2010ApJ...708.1025K}, and can account for the energetic mixing of radioactive $^{56}$Ni in the ejecta \citep{2017ApJ...845L..11H, 2018ApJ...852..100M}. 
The extra emission would appear as a "bump" feature in the early light curves, which contain valuable information about the mass-loss history of the progenitor system, but which will be inaccessible a few days later on account of the destruction of its ejecta \citep{2020ApJ...902...46Y}. 
The bump feature, which suggests that SNe Ia may originate from SD progenitor systems, has possibly been detected in some SNe Ia. Examples include SN 2012cg \citep{2016ApJ...820...92M, 2018ApJ...855....6S}, iPTF14atg with an ultraviolet (UV) excess \citep{2016MNRAS.459.4428K}, iPTF16abc \citep{2018ApJ...852..100M}, MUSSES1604d with an early red flux excess \citep{2017Natur.550...80J}, SN 2017cbv with a resolved blue bump \citep{2017ApJ...845L..11H, 2018ApJ...863...24S}, SN 2017fgc \citep{2021ApJ...919...49Z}, SN 2018aoz with a faint excess in the $V$ and $i$ bands but a plateau in the $B$ band during its infant phase \citep{2022NatAs.tmp...45N}, SN 2018byg with a bump that could result from an SD and interaction or a DD and Ni mixing \citep{2019ApJ...873L..18D}, SN 2018oh with a pronounced initial flux excess \citep{2019ApJ...870...12L, 2019ApJ...870...13S, 2019ApJ...870L...1D}, and SN 2019yvq with an early bright UV excess \citep{2021ApJ...919..142B}. 

In this article, we present the optical photometric and spectroscopic observations of Type Ia SN 2021hpr. Observations and data reductions are introduced in Section \ref{sec:obs and data reduc}. Section \ref{sec:phot prop} presents the optical light curves and the evolution of the spectra. The comparison of the light/color curves and spectra with those of other SN Ia, and the properties of the host galaxy are discussed in Section \ref{sec:discussion}. The conclusion is given in Section \ref{sec:conclusion}.

\section{Observations and Data Reductions} \label{sec:obs and data reduc}

\begin{figure}
\plotone{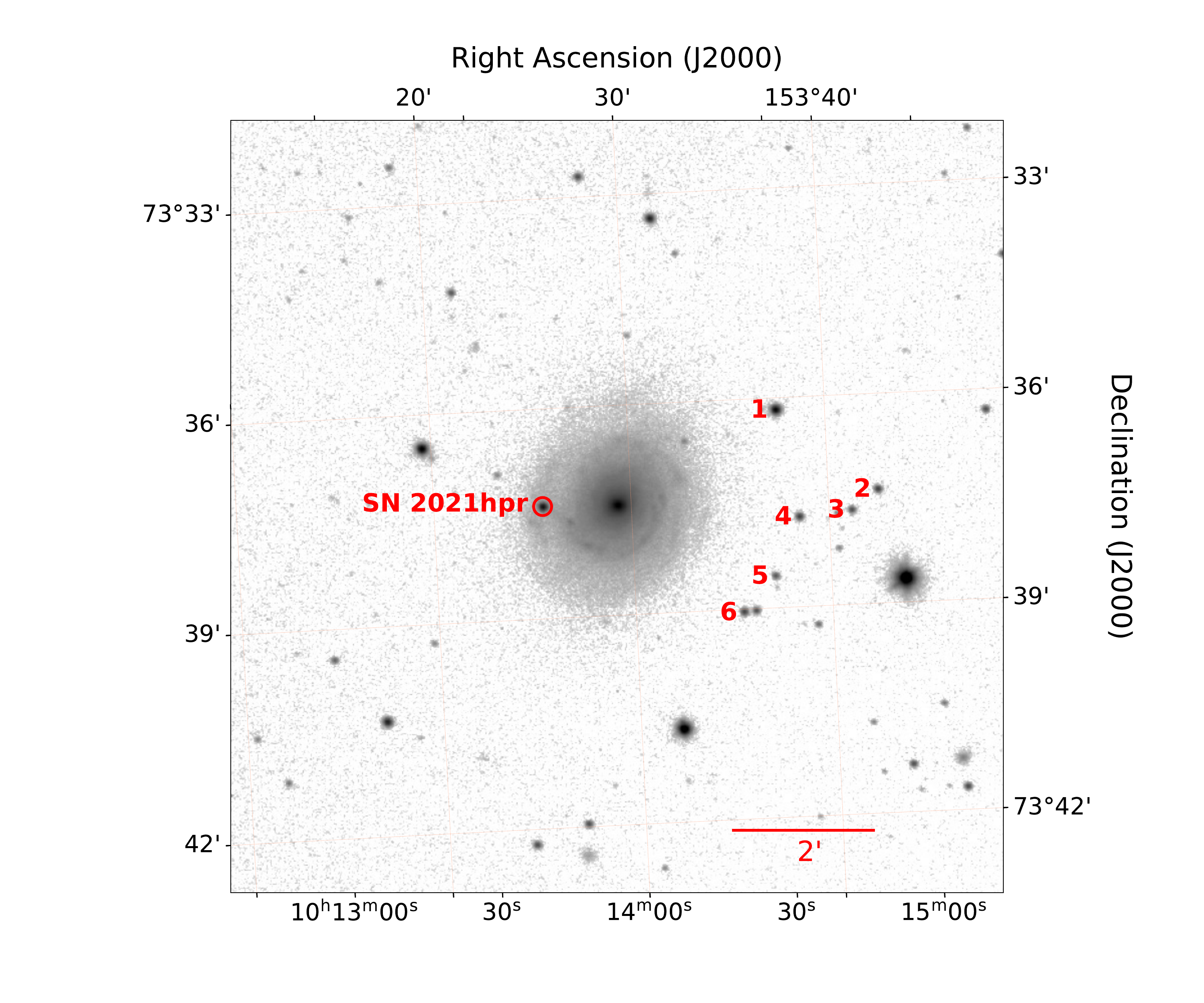}
\caption{The $R$-band image from the 60cm telescope at XingLong Observatory, NAOC, on 2021 April 18. SN 2021hpr is marked with a red circle, and its name is labeled nearby. Six reference stars are numbered on their left side. The bar corresponds to $2'$.\label{fig:SNlocation}}
\end{figure}

SN 2021hpr was discovered by Koichi Itagaki \citep{2021TNSTR.998....1I} at 10:46:28 on 2021 April 2 (UT dates are used throughout this paper) with an unfiltered magnitude of 17.7 mag. As shown in Figure \ref{fig:SNlocation}, SN 2021hpr is located at $-3'44''.28$ west and $-0'0''.54$ south of the nucleus of the host galaxy NGC 3147, and its 
coordinates are $\alpha = 10^{\text{h}}16^{\text{m}}38^{\text{s}}.680$, $\delta = +73^{\circ}24^{\prime}01^{\prime \prime}.80$ (J2000.0) \citep{2021TNSTR.998....1I}.
The $g$- and $r$-band light curves of SN 2021hpr (of which the first detection start from 2021 April 1) shown in the ALeRCE \citep{2021AJ....161..242F} explorer web interface \footnote{\url{alerce.online}} suggest that the luminosity of SN 2021hpr is rising rapidly. 
In addition, the ALeRCE stamp classifier \citep{2021AJ....161..242F} based on a convolutional neural network shows that SN 2021hpr has a high probability to be an SN. 
The first spectrum of SN 2021hpr taken at 19:48:36 on 2021 April 2 shows an expansion velocity of about 21,000 km s$^{-1}$ as measured from the broad Si {\sc ii} $\lambda 6355$ absorption, which is consistent with that of SNe Ia \citep{2021TNSCR1031....1T}. 

\subsection{Photometry}\label{sec:photometry}
The $BVRI$-band photometric observations of SN 2021hpr were carried out by using the 60cm telescope at XingLong Observatory, National Astronomical Observatories, Chinese Academy of Sciences (NAOC). The telescope is equipped with a 2048 $\times$ 2048 pixel CCD, giving a field of view (FOV) of $37.4' \times 37.4'$ (pixel scale $\sim 1^{\prime \prime}.1 \text{pixel}^{-1}$). 
The photometric data were obtained from 2021 April 3 (two days after discovery) to 2021 June 3, covering the peak phase and including 26 observing nights. The observations of SN 2021hpr followed the sequence of $B$, $V$, $R$, and $I$ filters and usually were repeated three times in each observation night.
Since SN 2021hpr is located in one spiral arm of its host galaxy, the contamination due to the galaxy should be removed carefully. The template images of the host galaxy were obtained about 240 days after the SN explosion, when the SN was faint.

The ccdproc package \citep{matt_craig_2021_5796852} was employed to do the bias subtraction and flat field correction of the SN images. Generally, the combined bias frames and combined twilight flat fields taken on the same night were adopted. 
Cosmic rays were removed using LACosmic \citep{2001PASP..113.1420V}. 
After these processes, astrometric calibrations of SN images were conducted by using Astrometry.net \citep{2012ascl.soft08001L}.
Then, the SN images observed each day with the same filter (usually three images) were combined using the reproject package \citep{2020ascl.soft11023R} and the ccdproc package. 
The template images obtained on 2021 December 24 were subtracted from the SN images after aligning. The point-spread functions (PSFs) of these two images were matched by adopting a Gaussian convolution, and the fluxes of the template images were scaled to the SN image by utilizing the fluxes of the same field stars in these two images.

Aperture photometry was performed on the template-subtracted images to acquire the instrumental magnitude of SN 2021hpr and of the local reference stars. 
These reference stars were used to calibrate the instrument magnitude to the standard Johnson $BV$ \citep{1966CoLPL...4...99J} and Kron$-$Cousins $RI$ \citep{1980SAAOC...1..234C} systems. 
The $BVRI$ magnitudes of the six reference stars (see Figure \ref{fig:SNlocation}) are given in Table 2 of \cite{2012A&A...537A..57B}. 
Here, we give the final photometry result of SN 2021hpr in Table \ref{tab:phot}.

The magnitudes errors take into account the Poisson noise of the target signal, the noise from image combination, the noise from the photometric calibration, and the noise of the sky background.

\subsection{Spectroscopy}\label{sec:spectroscopy}
The optical spectra of SN 2021hpr were obtained with the Beijing Faint Object Spectrograph and Camera (BFOSC) attached to the 2.16m telescope at XingLong Observatory, NAOC \citep{2016PASP..128k5005F}. 
A grism of G4 + 385LP with a wavelength range of $3700$ to $8800~\text{\AA}$ and a slit width of $2''.3$ was used for all the observations.
The spectral resolution is about 20 $\text{\AA}$, determined from the sky emission lines.

The spectroscopic observations were carried out from 2021 April 3 to June 20, covering the peak phase, and a total of six spectra were obtained. The spectra of the SN and a standard star were obtained at a similar airmass for flux calibration. 
The log of the spectroscopic observations is given in Table \ref{tab:spec}. 
The spectral data were reduced using standard IRAF routines, including bias subtraction, flat correction, and cosmic-rays removal. 
The SN 2021hpr spectra were wavelength calibrated using observations of an FeAr lamp taken in the dome on the same night. 
Additionally, a spectrum of the nucleus of NGC 3147 was observed on 2022 February 1 using BFOSC with the same grism and slit in order to analyze the host galaxy of SN 2021hpr, and it will be discussed in section \ref{sec:host galaxy}.

\section{Optical Light Curves and Spectra of SN 2021hpr} \label{sec:phot prop}

\subsection{Optical Light Curves}
\label{sec:light curve}

\begin{figure}
\plotone{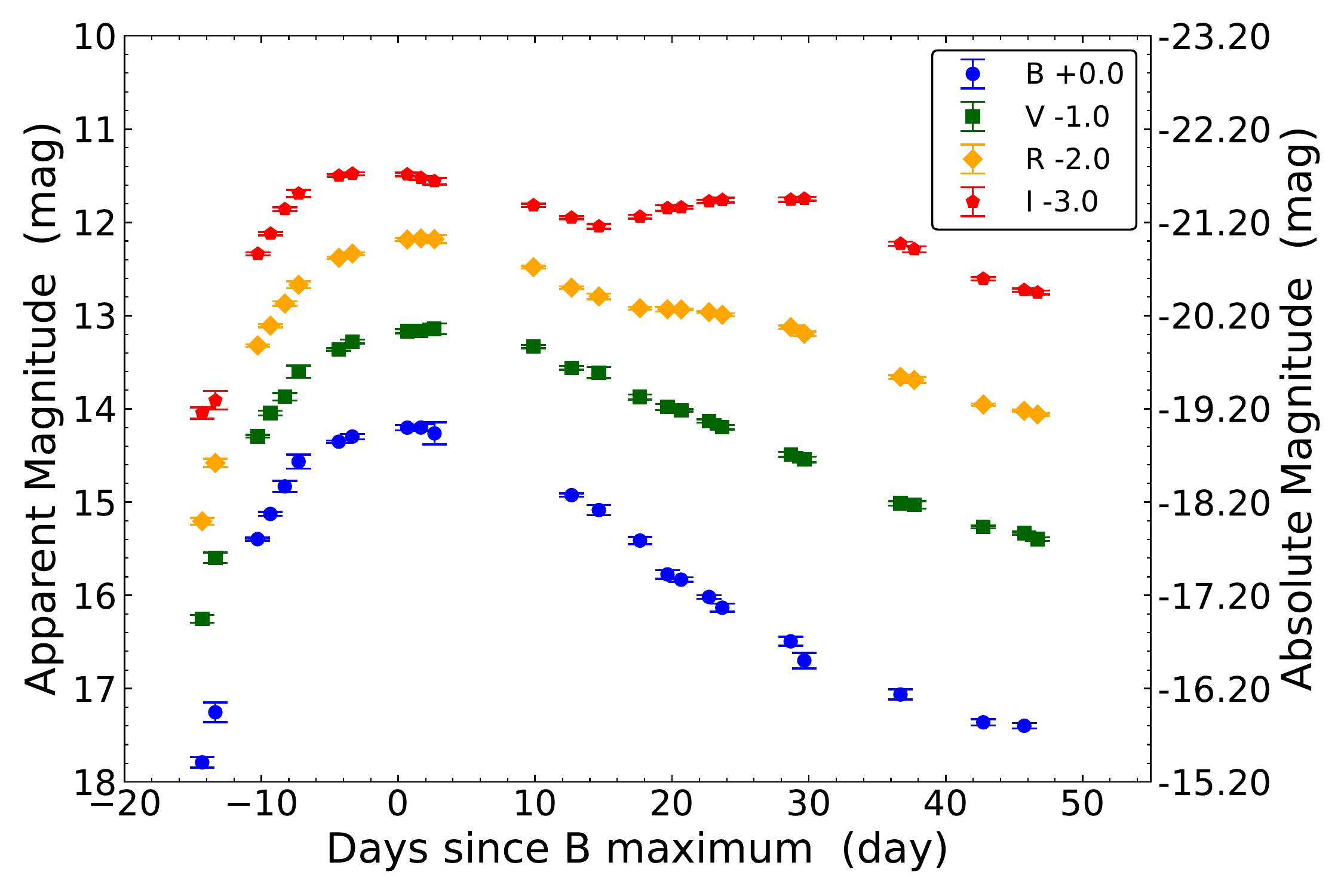}
\caption{$BVRI$ light curves of SN 2021hpr. The light curves are shifted by different constants in each band. The error bar corresponds to $1\sigma$ magnitude. \label{fig:light curve}}
\end{figure}

The optical light curves of SN 2021hpr in the $BVRI$ bands are shown in Figure \ref{fig:light curve}, covering the phases from $-$14.37 to +46.67 days relative to the $B$-band maximum. The photometric evolution of SN 2021hpr is consistent with that of typical normal SNe Ia. 
A plateau is evident in the $R$-band light curve from t $\sim$ +18 to $\sim$ +25 days relative to peak luminosity, while a secondary hump is seen in the $I$-band light curve. 
The $B$-band magnitudes around maximum light were fitted using a second-order polynomial to estimate the $B$-band peak magnitude, which is $14.017 \pm 0.017$ mag on MJD $59,321.862 \pm 0.450$. 
The $V$-band peak magnitude is $14.091\pm 0.014$ on MJD $59,323.220\pm 0.733$, which is about 1.4 days later. The peak magnitudes and corresponding MJDs of the $BVRI$ bands are calculated and listed in Table \ref{tab:mjd m in peak}.
The post-maximum magnitude decline rate $\Delta m_{15}(B)$; \citep{1993ApJ...413L.105P} is estimated by SNooPy \citep{2011AJ....141...19B} to be $0.949\pm 0.019$ mag. It is slightly lower than the typical value of 1.1 mag \citep{1999AJ....118.1766P}, but still in the range for SNe Ia of 0.75$-$1.94 \citep{2003AJ....125..166K}.

The rise time represents the time from explosion to maximum luminosity, and is dependent on the progenitors of SNe Ia \citep{1992ApJ...401...49L}. 
Assuming SN 2021hpr exploded like an expanding fireball, with a constant temperature and velocity, its luminosity or flux $f$ should be proportional to the second power of time \citep{1999AJ....118.2668R}. The rise time $t_{rise}$ can be obtained from the relation:
\begin{equation}
f(t) = \alpha(t+t_{rise})^2, \label{eq: rise time}
\end{equation}
where $t$ represents the phase relative to the $B$-band maximum  and $\alpha$ describes the velocity of the rising flux.
This relationship is consistent with the SNe light-curve patterns of some sky surveys \citep[i.e.,][]{1999AJ....118.2668R, 2001ApJ...558..359G, 2006AJ....132.1707C, 2007AJ....133..403G, 2010ApJ...712..350H, 2011MNRAS.416.2607G, 2015MNRAS.446.3895F}  and of some well-sampled individual SNe Ia \citep[i.e., SN 2011fe,][]{2011Natur.480..344N}.

Based on Equation (\ref{eq: rise time}), the rise time of SN 2021hpr can be derived as $16.424\pm 0.078$ days by fitting the data from the first observation day to $\sim -$8 days. 
The data taken on 2021 April 3 were not included for data fitting due to the large errors. 

\subsection{Evolution of the Spectra} \label{subsec:spec evo}

\begin{figure}
\includegraphics[angle=0,width=170mm]{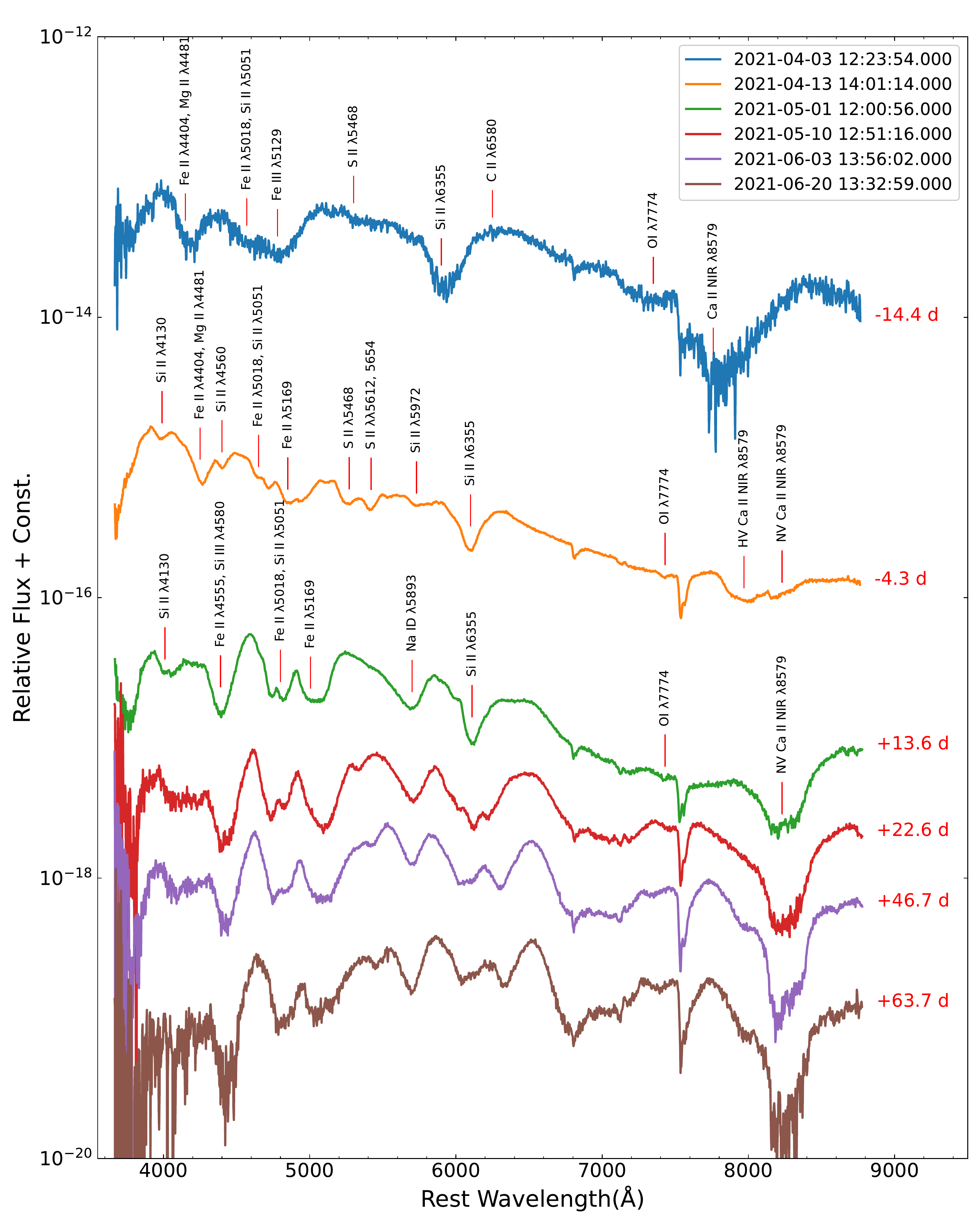}
\centering
\caption{Spectral sequence of SN 2021hpr from $t\sim -$14.4 days to $\sim$ +63.7 days relative to the $B$-band maximum, obtained with 2.16m telescope at XingLong Observatory, NAOC. The redshift of the host galaxy (z=0.009346) has been corrected for. The phase relative to the $B$-band maximum of each spectrum is denoted on the right side. \label{fig:spec evolution}}
\end{figure}

The spectral sequence of SN 2021hpr is shown in Figure \ref{fig:spec evolution}. A total of six optical low-resolution spectra cover the phases from $-$14.37 days to +63.68 days relative to the $B$-band maximum. 
The redshift of the host galaxy was corrected for in all spectra. The first spectrum of SN 2021hpr obtained at t $\sim -$14.37 days is the earliest one among all publicly available spectra. 

In the first spectrum of SN 2021hpr, the absorption lines are mainly those of intermediate-mass elements, such as Mg {\sc ii}, Si {\sc ii}, S {\sc ii}, and Ca {\sc ii}. 
The most distinct features characterized by Si {\sc ii} $\lambda$6355 and the Ca {\sc ii} NIR triplet (8498, 8542, and 8662 $\text{\AA}$) are broad and strong with a very large velocity, and they can be clearly identified near 5900 $\text{\AA}$ and 7800 $\text{\AA}$, respectively. 
The blended features of Fe {\sc ii} $\lambda$4404 and Mg {\sc ii} $\lambda$4481 can be found near 4200 $\text{\AA}$, and the broad absorption feature near 4700 $\text{\AA}$ could be caused by Fe {\sc ii} $\lambda$5018, Si {\sc ii} $\lambda$5051 and Fe {\sc iii} $\lambda$5129. 
The weak feature near 7400 $\text{\AA}$ should be O {\sc i} $\lambda$7774. 
No apparent Na I D doublet lines could be identified in the spectrum, which indicates that the interstellar extinction can be ignored. 

At $t\sim -4.31$ days, the characteristic of Si {\sc ii} $\lambda$4130 feature appeared in the spectrum, and the absorption features of Fe {\sc ii} $\lambda$4404, Mg {\sc ii} $\lambda$4481, Si {\sc iii} $\lambda$4560, Fe {\sc ii} $\lambda$5018, Si {\sc ii} $\lambda$5051 and Fe {\sc iii} $\lambda$5129 became more obvious and more recognizable. 
The "W" shape consisting of S {\sc ii} $\lambda$5468 and S {\sc ii} $\lambda \lambda$5612,5654 emerged. 
Si {\sc ii} $\lambda$5972 should be responsible for the absorption line near 5700 $\text{\AA}$. 
The expansion velocity of Si {\sc ii} $\lambda$6355 decreased rapidly. The HV component can be distinguished from the NV component of the Ca {\sc ii} NIR triplet, of which the velocity became smaller.

Two weeks after the maximum luminosity, at $t \sim 13.6$ days, the absorption feature of Si {\sc iii} $\lambda$4560 became weaker, and the "W" shape almost disappeared. 
Near 5700 $\text{\AA}$, a strong absorption feature could be identified as Na I D $\lambda$5893, while the absorption feature of Si {\sc ii} $\lambda$5972 was barely visible.
The expansion velocity of the Ca {\sc ii} NIR triplet decreased further. 

From the third week to the second month since peak luminosity, according to the spectra observed at t $\sim$ 22.6, $\sim$ 46.7, and $\sim$ 63.7 days, the iron-peak element features became stronger and gradually dominated the spectrum. 
The absorption feature of Si {\sc ii} $\lambda$6355 became weaker, while the Fe {\sc ii} lines gradually appeared.

\section{Analysis and Discussion} \label{sec:discussion}

\subsection{Light-Curve Comparison}\label{sec:light curve compare} 

\begin{figure}
\plotone{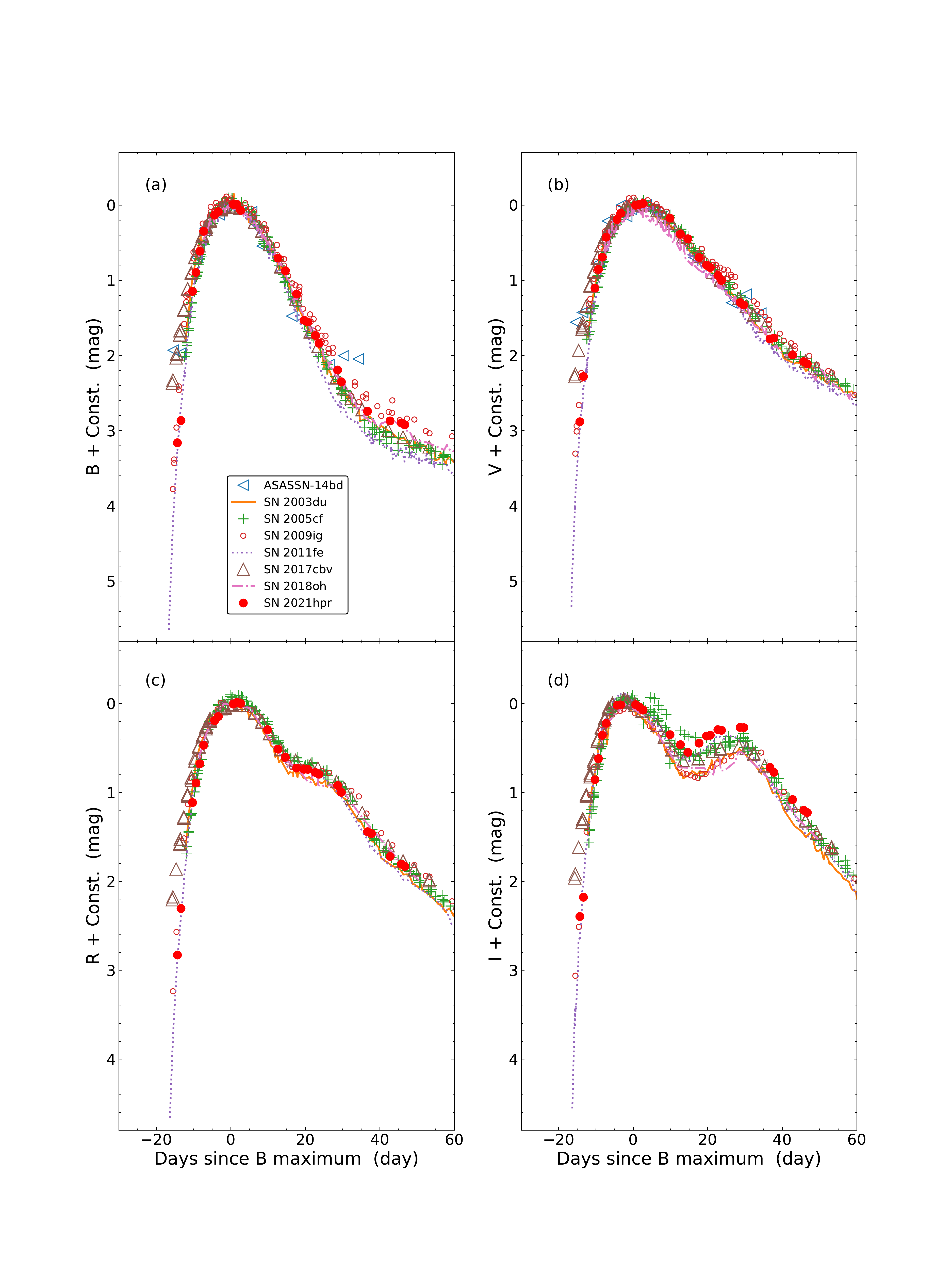}
\caption{Comparisons of the $BVRI$ photometric light curves of SN 2021hpr with those of SNe 2003du, 2005cf, 2009ig, 2011fe, 2017cbv, 2018oh, and ADSSSN-14bd (see the text for references). The light curves of all SNe are shifted according to their the $B$-band maximum time $t_{\text{max}}^{\text{B}}$ and the $B$-band maximum magnitude $m_{\text{max}}^{\text{B}}$ to match the peak in all bands.\label{fig:light curve compare}}
\end{figure}

In Figure \ref{fig:light curve compare}, the $BVRI$-band light curves of SN 2021hpr are plotted with those of other SNe Ia, which are ASASSN-14bd \citep{2014Ap&SS.354...89B}, SN 2017cbv \citep{2020ApJ...904...14W}, SN 2003du \citep{2005A&A...429..667A, 2007A&A...469..645S, 2009ApJ...700..331H, 2010ApJS..190..418G, 2012MNRAS.425.1789S}, SN 2005cf \citep{2007MNRAS.376.1301P, 2009ApJ...700..331H, 2010ApJS..190..418G, 2012MNRAS.425.1789S, 2014Ap&SS.354...89B}, SN 2009ig \citep{2012ApJ...749...18B, 2012ApJS..200...12H, 2012MNRAS.425.1789S, 2014Ap&SS.354...89B, 2019MNRAS.490.3882S}, SN 2011fe \citep{2012JAVSO..40..872R, 2012MNRAS.425.1789S, 2013CoSka..43...94T, 2013NewA...20...30M, 2014Ap&SS.354...89B, 2019MNRAS.490.3882S}, and SN 2018oh \citep{2019ApJ...870...12L}. The first two SNe Ia have a flux excess in their early phase, while the others are normal SNe Ia. 
The light curves of all samples are normalized to their peaks. 
It can be seen that the $B$-band light curve of SN 2021hpr has a similar shape to that of other normal SNe Ia in the phase before the maximum luminosity. 
The decline rate $\Delta m_{15}(B)$ of SN 2021hpr (see Section \ref{sec:light curve}) is close to the $0.96 \pm 0.02$ mag of SN 2008oh \citep{2019ApJ...870...12L}, $0.89 \pm 0.02$ mag of SN 2009ig \citep{2012ApJ...744...38F}, and $0.990 \pm 0.013$ mag of SN 2017cbv \citep{2020ApJ...904...14W}. The decline rate is slower than that of SN 2011fe, which has a typical value for the decline rate of $1.103 \pm 0.035$ mag, while its luminosity is brighter. 
After $t\sim$ +30 days, the $B$-band luminosity is relatively brighter than those of other SNe Ia during the same phase. 
The $VR$-band light curves of SN 2021hpr follow the evolution of other normal SNe Ia. 
The second maximum in the $I$ band of SN 2021hpr is slightly stronger than those of other SNe Ia. 
Furthermore, ASASSN-14bd (in $BV$-band) and SN 2017cbv are brighter than the others before reaching the maximum.

\subsection{The Reddening and Color Evolution}\label{sec:reddening and color evolution}
\begin{figure}
\includegraphics[angle=0,width=100mm]{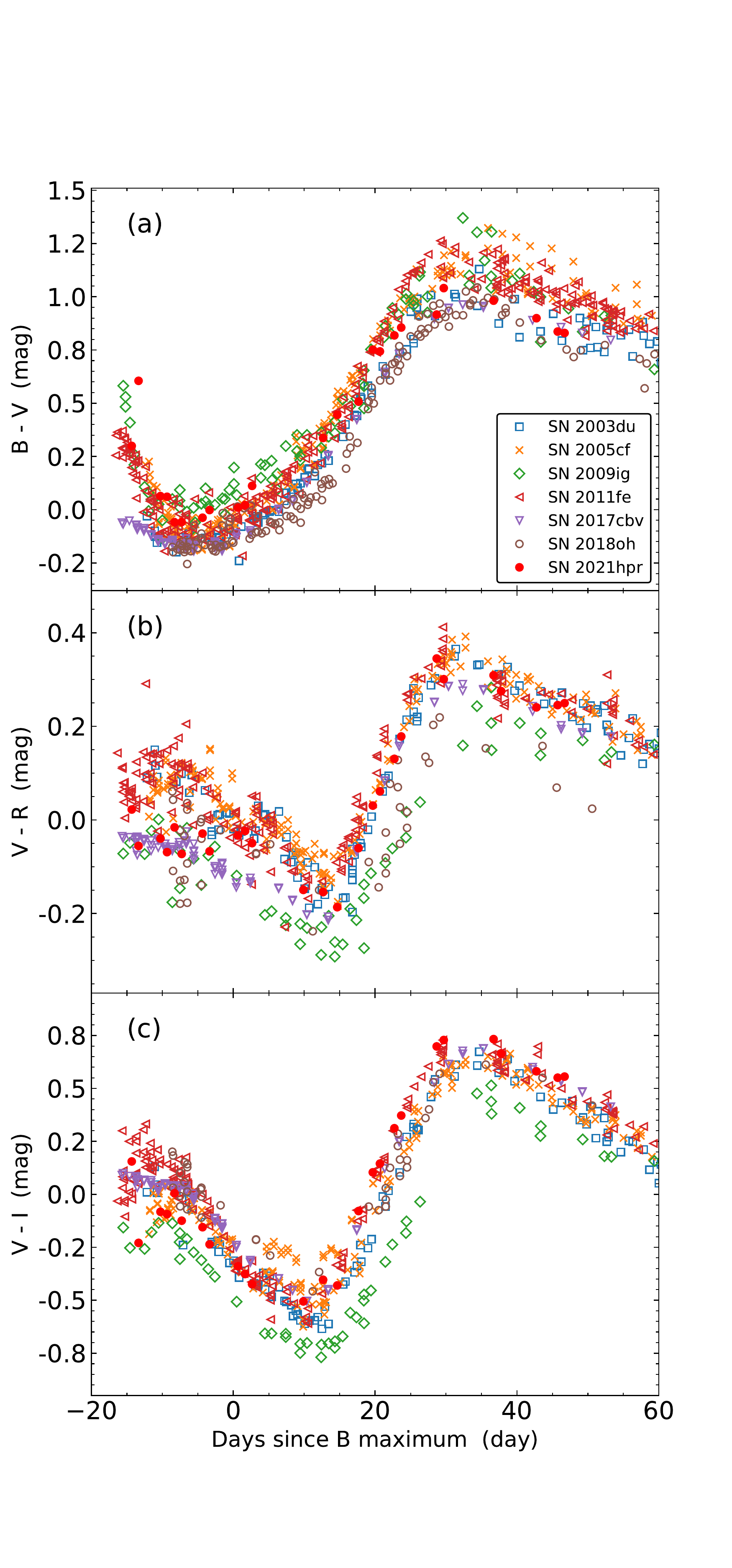}
\centering
\caption{The $B-V$, $V-R$, and $V-I$ color curves of SN 2021hpr compared with those of normal type Ia SNe 2003du, 2005cf, 2009ig, 2011fe, and 2018oh (see the text for the references). All SNe have been dereddened. \label{fig:color curve compare}}
\end{figure}

In Figure \ref{fig:color curve compare}, $B-V$, $V-R$, and $V-I$ color curves of SN 2021hpr are compared with those of the other SNe Ia in Section \ref{sec:light curve compare}, 
except for ASASSN-14bd because it lacks data in the $RI$-band. All the color curves have been corrected for the Galactic and host galaxy reddening. 
As for SN 2021hpr, the Galactic extinction is estimated to be $A_B = 0.088$ mag, $A_V = 0.067$ mag, $A_R = 0.053$ mag and $A_I = 0.037$ mag on the basis of the dust map of the Milky Way \citep{2011ApJ...737..103S}. 
The host galaxy extinction of SN 2021hpr is almost negligible, which will be discussed in Section \ref{sec:host galaxy}.

From $t\sim -15$ to +20 days, the $B-V$ color evolution of SN 2021hpr shows a resemblance to most of the selected SNe, while after $t\sim$ +20 days the color is bluer than most of the other supernovae. 
SN 2021hpr reaches the bluest $B-V$ color of $-$0.06 mag at $t\sim -5$ days, then gradually turns into its reddest color of $\sim 1.0$ mag at $t\sim +30$ days. 
SN 2017cbv is an SN Ia with early excess. Its $B-V$ color before the $B$-band maximum is bluer than that of most of the normal SNe, including SN 2021hpr, which suggests that SN 2021hpr may not have a bump feature in early phase.
Before reaching maximum luminosity, the $V-R$ color curve of SN 2021hpr has a plateau phase at $\sim 0.25$ mag, which is similar to that of SN 2017cbv. However, the latter starts to decline after t $\sim -7$ days. 
The $V-R$ color of SN 2021hpr reaches the minimum value of approximately $-0.2$ mag, and undergoes rapid reddening until it reaches the reddest color of $\sim 0.33$ mag at $t\sim +30$ days. 
The $V-I$ color of SN 2021hpr is similar to that of SN 2011fe. SN 2021hpr reaches its bluest $V-I$ ($\sim -0.5$ mag) at $t\sim +10$ days and its reddest $V-I$ ($\sim 0.8$ mag) at $t\sim +30$ days. In addition, the $V-I$ color of SN 2021hpr is redder than that of most of the selected SNe after $t\sim 20$ days. 

\subsection{Spectra Comparison and Spectroscopic Properties}\label{sec:spec compare}
\begin{figure}
\includegraphics[angle=0,width=180mm]{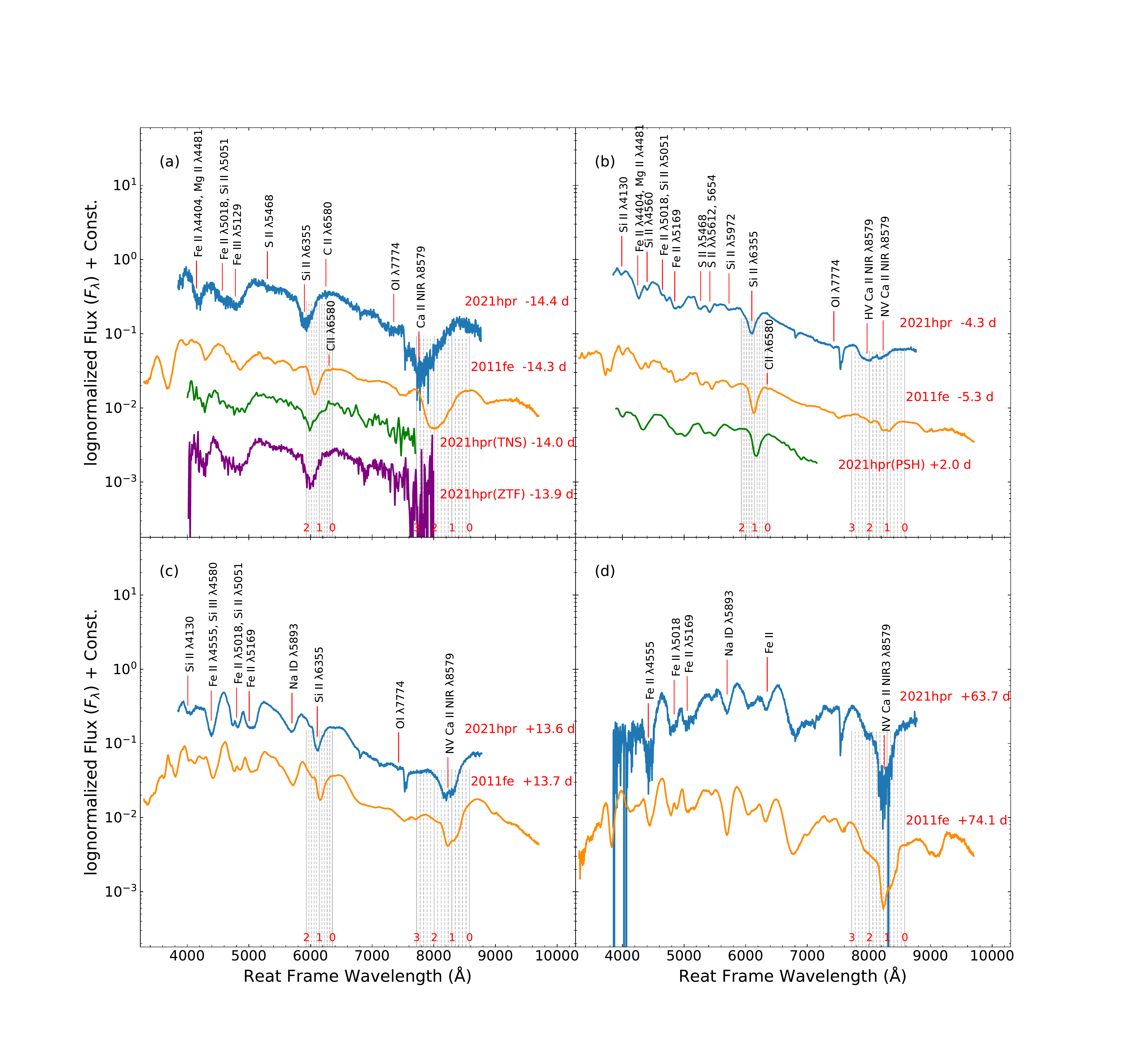}
\centering
\caption{Spectral comparisons between SN 2021hpr and SN 2011fe at four different phases. The spectra of SN 2021hpr in blue are obtained from the XingLong 2.16m telescope, and the spectra of SN 2021hpr in green or purple are obtained from the Transient Name Server. The spectra of SN 2011fe in yellow are acquired from \cite{2013A&A...554A..27P}. 
The redshift of the host galaxy has been corrected for. The target name and the observation time relative to the $B$-band maximum are labeled near each spectrum. The solid gray lines label the positions of blueshifted of velocities of $-$30,000, $-$20,000, $-$10,000, and 0 km s$^{-1}$ for Si {\sc ii} $\lambda$6355 and the Ca {\sc ii} NIR triplet and are shown as 3, 2, 1, and 0 in red. 
Intervals of 2000 km s$^{-1}$ are shown as dashed gray lines between the solid gray lines. \label{fig:spec compare}}
\end{figure}

In Figure \ref{fig:spec compare}, we compare the spectra of SN 2021hpr with that of 2011fe at four different epochs (t $\approx$ $-$14, $-$4, +13 and +70 days relative to the $B$-band maximum). All the spectra have been corrected for redshift. 
In addition, the spectra of SN 2021hpr before or near peak luminosity downloaded from the Transient Name Server (TNS) website are also added into comparison, and are shown in Figure \ref{fig:spec compare}a and Figure \ref{fig:spec compare}b. 
Furthermore, we obtained the velocity of Si {\sc ii} $\lambda$6355 of SN 2021hpr by Gaussian fitting. 

At t $\sim -14$ days, the spectra of SN 2021hpr obtained by 2.16 m telescope are similar to the spectra from TNS, but shifted nearly half a day earlier (see Figure \ref{fig:spec compare}a). 
The expansion velocity of Si {\sc ii} $\lambda$6355 of the former reaches $\sim$ 20,440 km s$^{-1}$, which is slightly higher than that in the latter spectrum from TNS. 
SN 2021hpr shows a very high velocity for the Ca {\sc ii} NIR triplet, which is nearly 28,000 km s$^{-1}$. 
The velocities of these two features are much higher than those of SN 2011fe at this phase. 
A weak absorption feature of C {\sc ii} $\lambda$6580 is seen on the redder side of Si {\sc ii} $\lambda$6355 in the spectrum of SN 2011fe, while no obvious carbon feature can be recognized in the spectrum of SN 2021hpr. At least 30$\%$ of SNe would show lines of carbon in their spectra before the luminosity reaches maximum \citep{2011ApJ...743...27T}.

About four days before maximum (Figure \ref{fig:spec compare}b), the Si {\sc ii} $\lambda$6355 velocity of SN 2021hpr decreased more rapidly to $\sim$ 12,420 km s$^{-1}$. The velocity gradient from $t\sim -14.37$ to $-4.31$ days is about 800 km s$^{-1}$ day$^{-1}$, which confirms that SN 2021hpr belongs to the HVG SNe Ia group. Even so, this velocity is still higher than that of SN 2011fe and other NV SNe Ia, whose velocity is about 11,800 km s$^{-1}$ \citep{2009ApJ...699L.139W}. 
This indicates that SN 2021hpr should be a transitional object between the HV and Normal groups, which is also confirmed by the slower $B$-band decline rate (see \ref{sec:light curve}). 
An HV component of the Ca {\sc ii} NIR triplet absorption profile can be found in the spectrum of SN 2021hpr, which is similar to that of SN 2011fe. 
The velocity range of this absorption feature is between 14,000 and 20,000 km s$^{-1}$, which is higher than that of SN 2011fe. The blended feature composed of Fe {\sc ii} $\lambda$4404 and Mg {\sc ii} $\lambda$4481 seems stronger in the spectrum of SN 201hpr than that in that of SN 2011fe. 

Figure \ref{fig:spec compare}c displays the spectra two weeks after maximum luminosity. The Si {\sc ii} and Ca {\sc ii} absorption features still remain, while the "W" shape of S {\sc ii} has almost disappeared. The Si {\sc ii} $\lambda$6355 velocity of SN 2021hpr is also slightly higher than that of SN 2011fe.
Figure \ref{fig:spec compare}d shows the spectra two months after maximum luminosity. 
SN 2021hpr and SN 2011fe show comparable absorption features in their spectra, and the Fe features became dominant.  
The Ca {\sc ii} NIR triplet absorption feature is still very strong in both spectra.

\subsection{Host Galaxy}\label{sec:host galaxy}

NGC 3147 is a face-on spiral galaxy and classified as SA(rs)bc \citep{2013AJ....145..138B} with a redshift of 0.00935 and a size of about $3'.9 \times 3'.4$. Its apparent magnitude in the $V$ band is 10.61 mag \citep{2007ApJS..173..185G}. 
\cite{2002A&A...394..435P} classified it as a type II Seyfert galaxy. 
However, a recent study discovered very broad emission lines in its nucleus \citep{2019MNRAS.488L...1B}, suggesting that it is a low accretion rate type I active galactic nucleus with a massive black hole.

\begin{figure}
\includegraphics[angle=0, width=140mm]{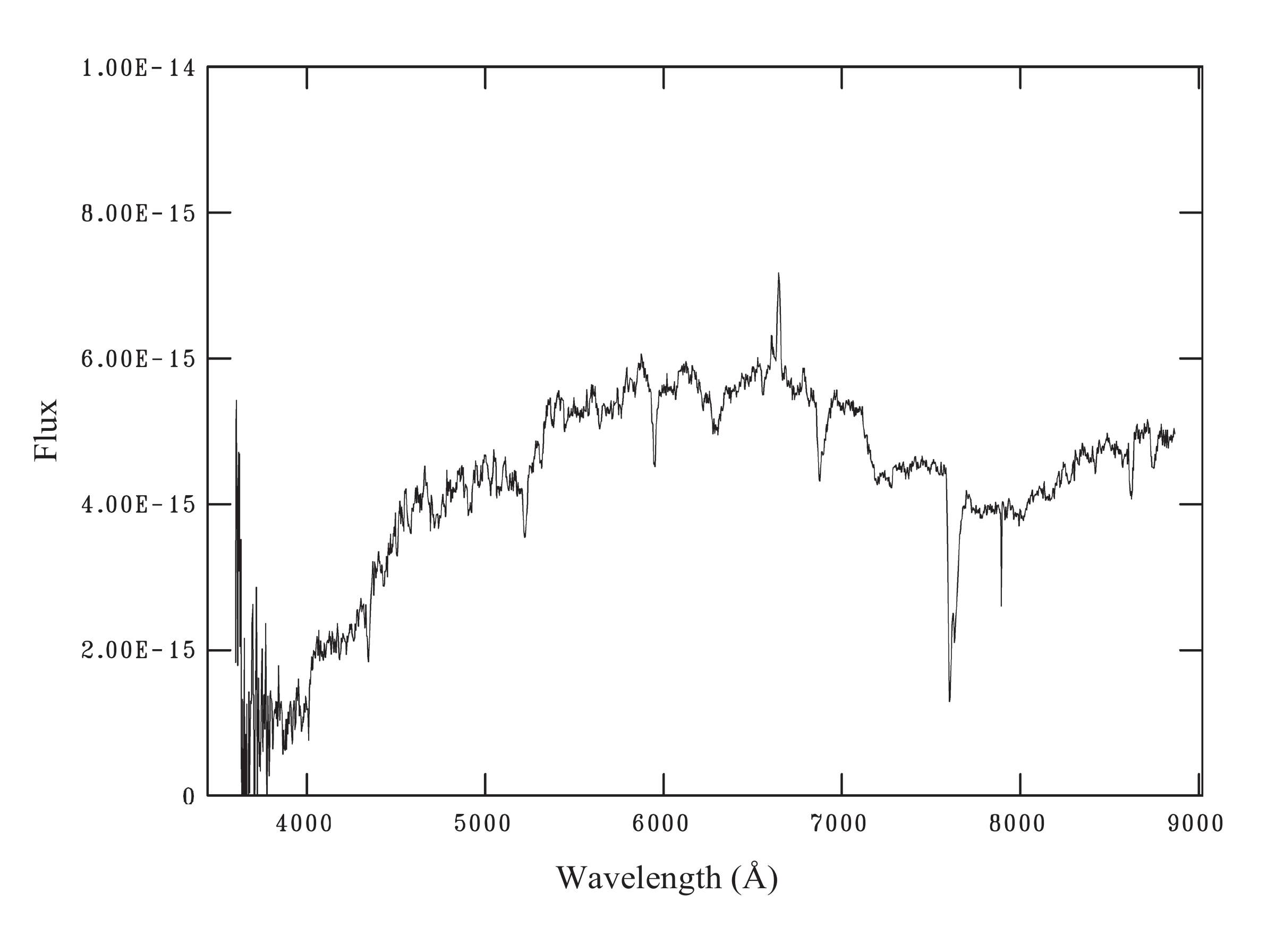}
\centering
\caption{The spectrum of the nucleus of the host galaxy (NGC 3147) of SN 2021hpr. The spectrum was obtained with the 2.16m telescope at XingLong Observatory, NAOC, on 2022 February 1. \label{fig:ngc3147spec}}
\end{figure}

\begin{figure}
\includegraphics[angle=0,width=140mm]{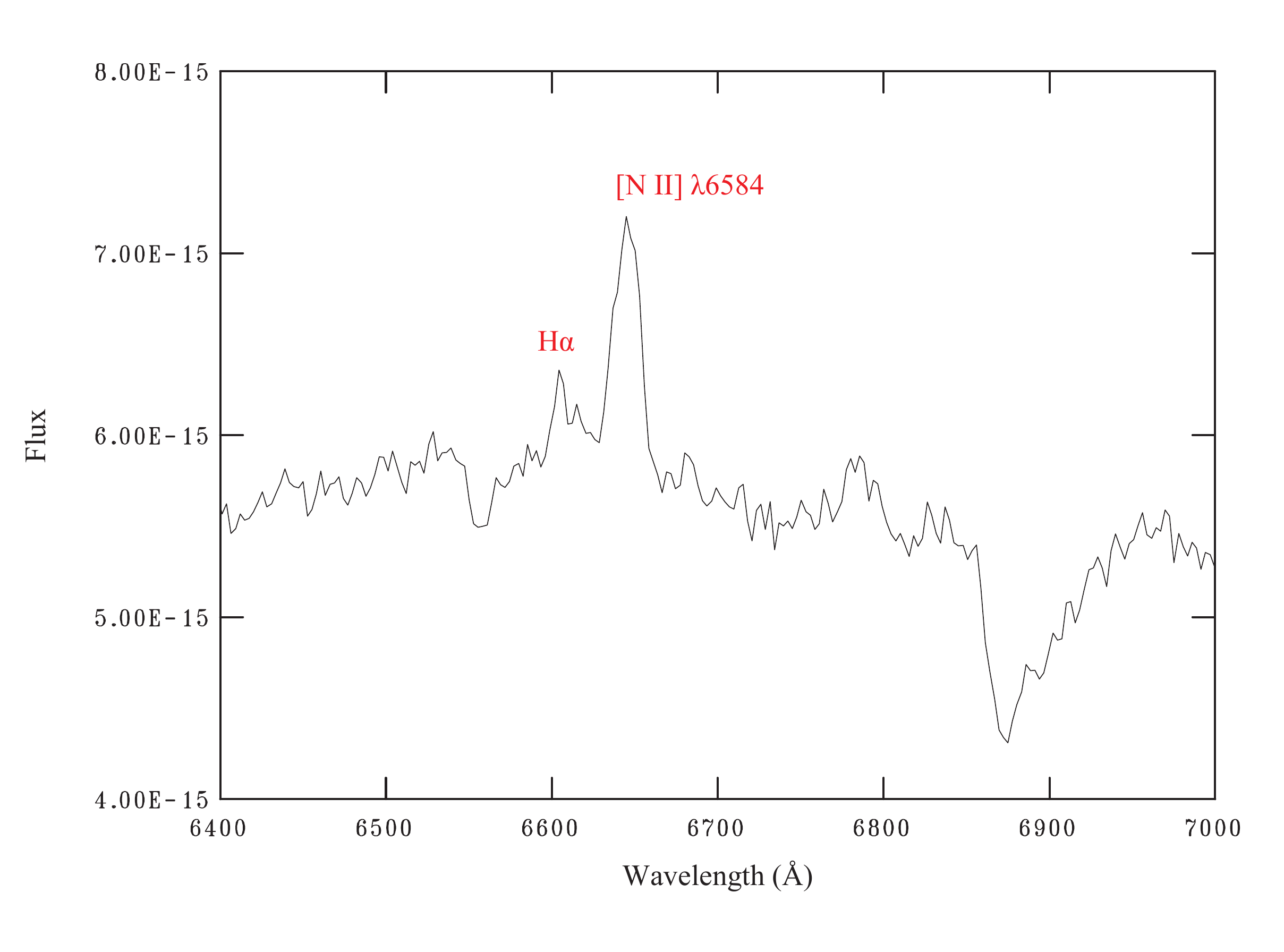}
\centering
\caption{A portion of the spectrum of the nucleus of the host galaxy (NGC 3147) shown in Figure \ref{fig:ngc3147spec}. 
The emission features of H$\alpha$ and [N {\sc ii}] $\lambda$6584 are labeled with red text nearby.
\label{fig:ngc3147ems}}
\end{figure}

We observed the nucleus of NGC 3147 with BFOSC on 2022 February 1, as shown in Figure \ref{fig:ngc3147spec}. 
The H$\alpha$ region of the spectrum is shown in Figure \ref{fig:ngc3147ems}. 
The ratio of H$\alpha$ and [N {\sc ii}] $\lambda$6584 suggests that NGC 3147 could be classified as a low-ionization nuclear emission-line region (LINER).

\begin{figure}
\includegraphics[angle=0,width=140mm]{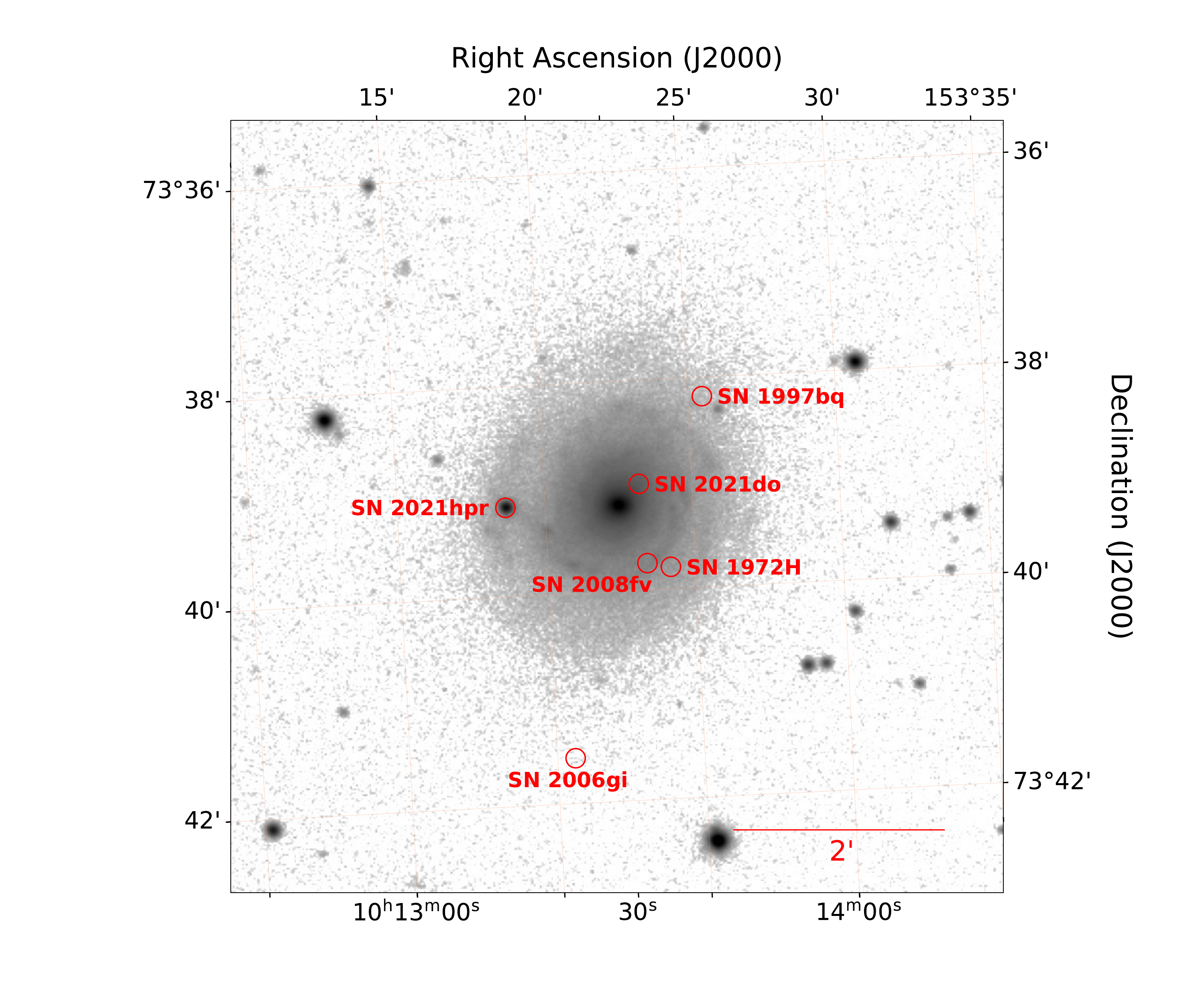}
\centering
\caption{The locations of SN 1972H \citep{1997A&A...317..423P}, SN 1997 \citep{2006AJ....131..527J}, SN 2006gi \citep{2006IAUC.8755....3D}, SN 2008fv \citep{2010PZ.....30....2T}, SN 2021do, and SN 2021hpr in NGC 3147. All these supernovae are circled with red and their names are given nearby. The bar corresponds to $2'$. 
\label{fig:SNinNGC3147}}
\end{figure}

NGC 3147 is also an SNe factory in which six SNe have been observed over the past 50 yr (1972-2022): SN 1972H \citep{1997A&A...317..423P}, SN 1997 \citep{2006AJ....131..527J}, SN 2006gi \citep{2006IAUC.8755....3D}, SN 2008fv \citep{2010PZ.....30....2T}, SN 2021do, and SN 2021hpr. 
Four of them are SNe Ia, one is an SN Ib, and one is an SN Ic. 
In Table \ref{tab:SNe in NGC 3147}, we summarize the SN type and equatorial coordinates (J2000) of these SNe, whose locations are shown in Figure \ref{fig:SNinNGC3147}. 
Interestingly, the four SNe Ia are all located in the galactic disk, while the Type Ic SN 2021do is closest to the galactic nucleus, and the Type Ib SN 2006gi is the farthest from the galactic center, in the outskirt of the galactic disk.

According to the statistics of \cite{2009ApJ...698.1307T}, there are more than 16 galaxies in which SNe have been detected more than four times. 
The highest record is still held by NGC 6946, where 10 SNe have been observed in recent decades. It is followed by NGC 4303 with eight SNe, NGC 3690 with seven SNe, and then NGC 3147 with six SNe. 
The SN explosion rate in NGC 3147 is high if compared to the average of about three per galaxy per century. 
Using radio data, \cite{2009ApJ...698.1307T} estimates the SN explosion rate of NGC 3147 as 0.18 SNe per year, with 0.06 per year for Type Ia, 0.09 per year for Type Ibc, and 0.03 per year for Type II. 
On average, three SNe Ia, 4.5 SNe Ibc, and 1.5 SNe II will explode in NGC 3147 every 50 yr. 
Therefore, the observed explosion rate is slightly higher than that expected for SNe Ia in NGC 3147, but much lower for SNe Ibc and SNe II. 

\begin{figure}
\includegraphics[angle=0,width=140mm]{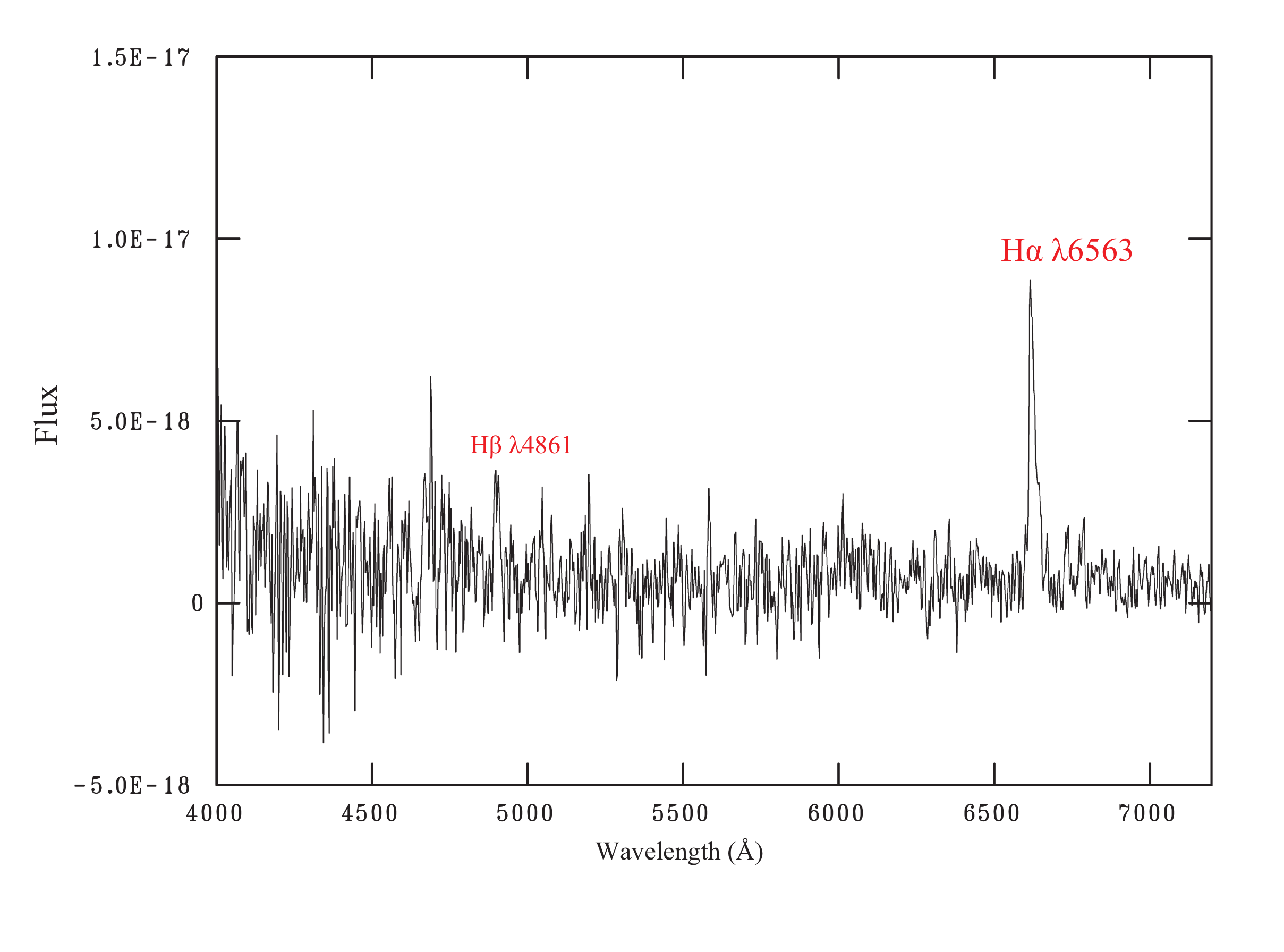}
\centering
\caption{The spectrum of a region nearby SN 2021hpr, obtained with the 2.16m telescope at XingLong Observatory, NAOC. The emission features of H$\alpha$ and H$\beta$ are labeled in red.
\label{fig:snnearspec}}
\end{figure}

\begin{figure}
\includegraphics[angle=0,width=140mm]{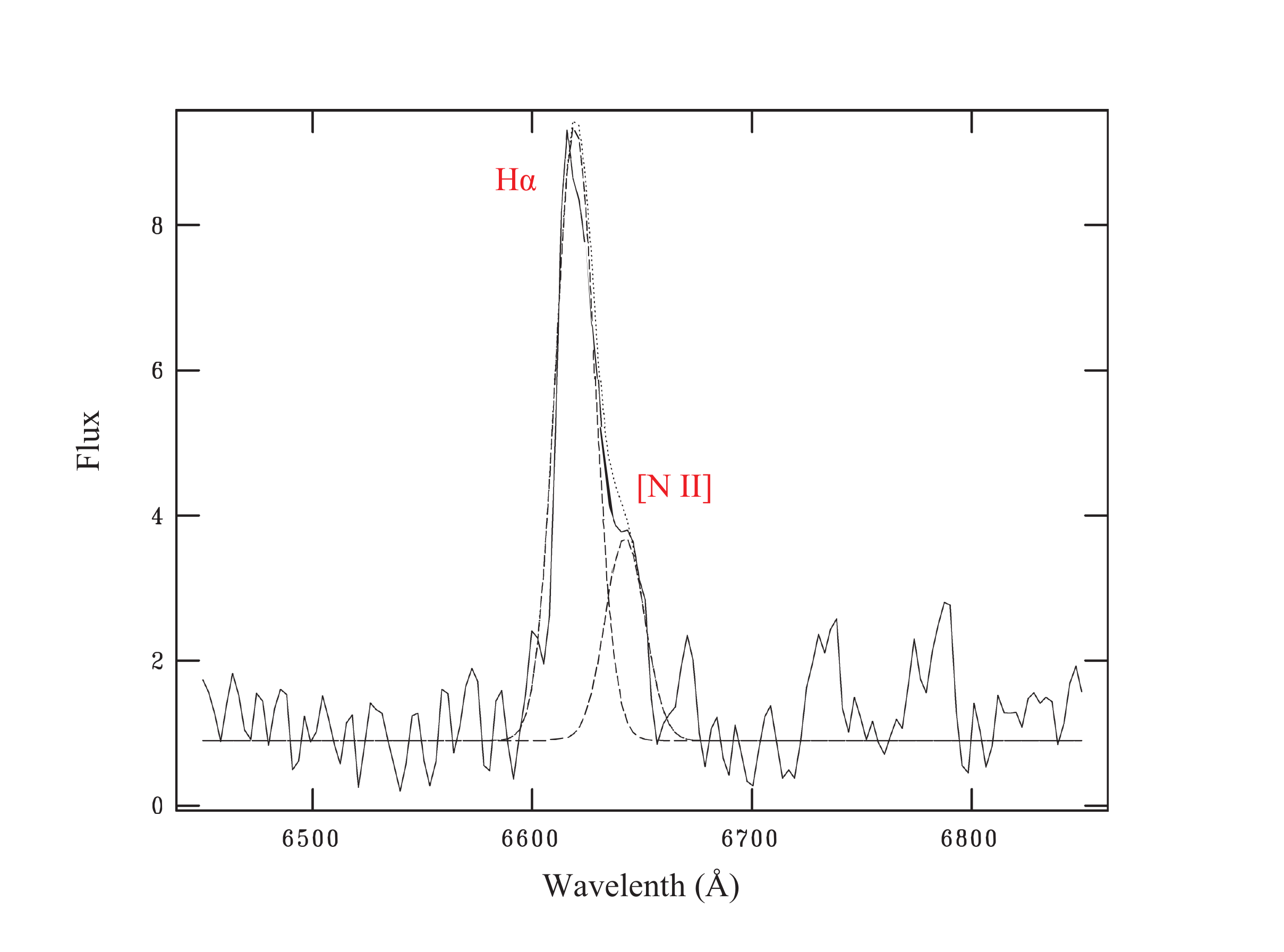}
\centering
\caption{The Gaussian fitting of the spectrum of the region nearby SN 2021hpr. Two gaussian components are labeled in red.
\label{fig:snfit}}
\end{figure}

Furthermore, we examined the metallicity of the surroundings of SN 2021hpr. 
Figure \ref{fig:snnearspec} shows the spectrum combined with spectra of three star-formation regions within the distance of SN 2021hpr of 30$^{\prime \prime}$. 
The flux ratio of decomposed H$\alpha$ and H$\beta$ is estimated to be $\sim 2.74$. Therefore, the host galaxy extinction of SN 2021hpr is negligible. 
The spectrum near 6600 $\text{\AA}$ displayed in Figure \ref{fig:snnearspec} is shown in Figure \ref{fig:snfit}, where it exhibits features of [N {\sc ii}] $\lambda$6584 and H$\alpha$ $\lambda$6563. 
Since the two emission lines are blended, we use double Gaussian fitting to derive the flux ratio as [N {\sc ii}] $\lambda$6584/H$\alpha$ $\lambda$6563 $\approx$ 0.33, corresponding to the metallicity of 12+log(O/H) $\approx$ 8.648, comparable to the solar metallicity of 8.69 \citep{2021A&A...653A.141A}. 
This is consistent with the suggestion by \cite{2020ApJ...895L...5P} that HV SNe might be present in massive galaxies with metal-rich environments. 

Based on the redshift of NGC 3147, 
we can obtain the the distance modulus of 
$m-M = 33.31~mag$.\footnote{Given by the NASA/IPAC Extragalactic Database \citep{https://doi.org/10.26132/ned3}, 
assuming $H_0$ = 67.8 km s $^{-1}$ Mpc$^{-1}$ 
(Virgo Infall only) and $\Omega_{\text{matter}}=0.308$} 
However, in the local universe, measurements need a better “standard
candle,” such as SNe Ia.
The distance modulus of NGC 3147 can be calculated if we know the absolute magnitude $M_{B,max}$ of SN 2021hpr. 
Based on the decline rate $\Delta m_{15}(B) = 0.949 \pm 0.019$ mag (see Section \ref{sec:light curve}) 
and the Phillips relation (\citealt{1999AJ....118.1766P}; e.g., \citealt{1995A&A...294L...9W}),
we can derive $M_{B,max} = -19.531\pm 0.210$ mag. 
Then, the distance modulus should be $m-M$ = $33.46\pm 0.21$ mag, which is the same as the result of \cite{2007ApJ...659..122J} ($m-M$ = $33.46\pm 0.11$ mag), who measured the distance modulus for SN 1997bq using the multicolor light-curve shape method (MLCS2K2). 
However, the distance moduli vary from 32.41 mag to 33.46 mag even when measured from the same SNe \citep[e.g. SN 1997bq;][]{2005ApJ...624..532R, 2006ApJ...645..488W, 2006ApJ...647..501P, 2007ApJ...659..122J, 2008ApJ...686..749K, 2008MNRAS.389.1577T, 2010ApJ...716..712A, 2013AJ....146...86T}.
While differences in the final result could be due to different physics, distance modulus measurements can be also affected by many other factors, such as photometric accuracy, internal reddening of the host galaxy, and template subtraction. 

\begin{figure}
\includegraphics[angle=0,width=140mm]{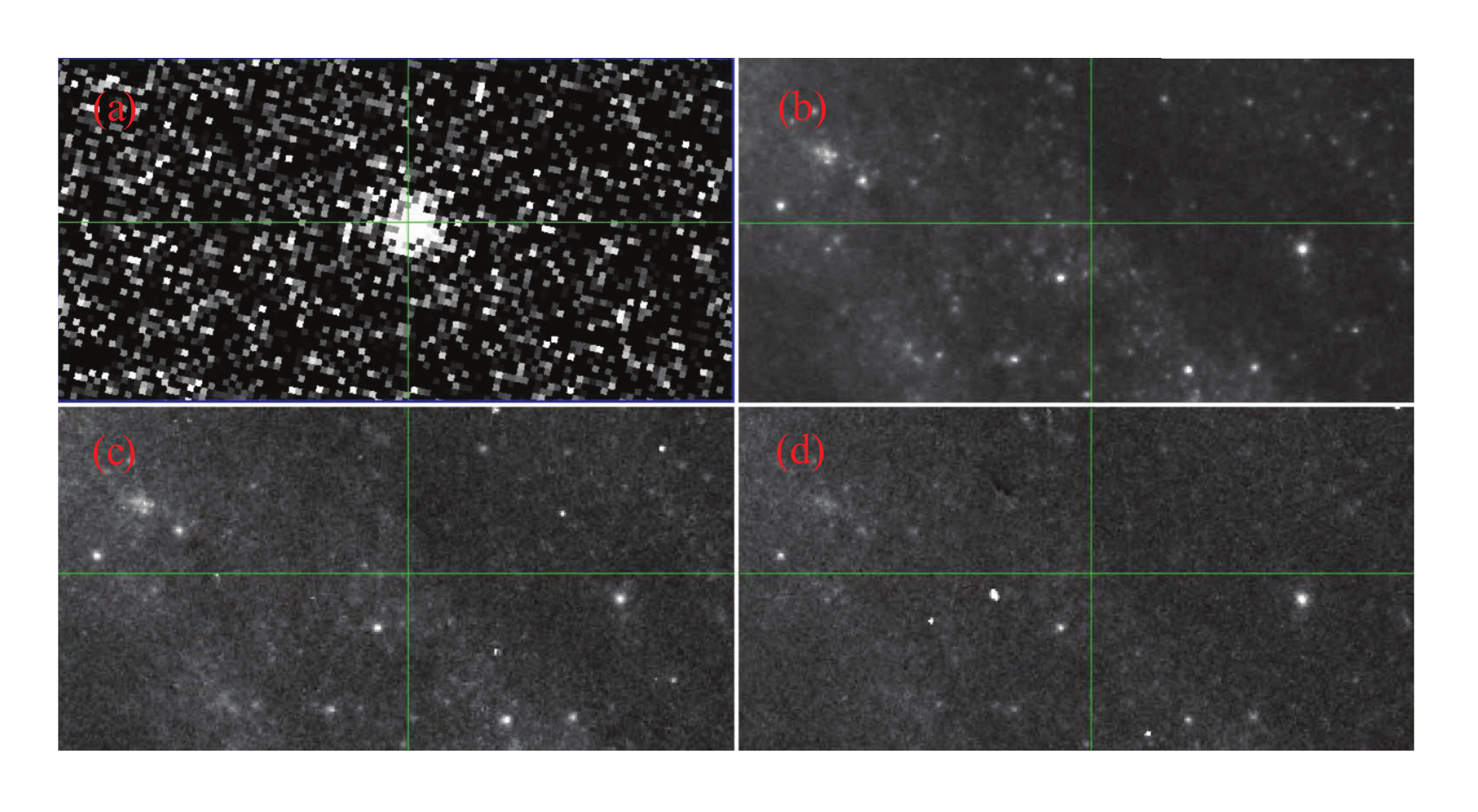}
\centering
\caption{The HST images in the area of SN 2021hpr. (a) The image in the WFC3/F125W band observed on 2021 April 29. SN 2021hpr appears at the position of the green cross. (b)-(d) The images in the WFC3/F350LP, WFC3/F555W, and WFC3/F814W bands observed on 2018 March 21, before SN 2021hpr exploded.
\label{fig:find progenitor with HST}}
\end{figure}

To search for the progenitor of SN 2021hpr, we inspected the Hubble Space Telescope (HST) images before and after (see Figure \ref{fig:find progenitor with HST}) SN 2021hpr exploded, and we failed to detect it. 
However, we can obtain an upper limit (3$\sigma$) for it of 28.50 mag, 28.35 mag and 27.87 mag (AB) in the WFC3/F350LP, WFC3/F555W and WFC3/F814W bands, respectively. 
Using the distance modulus obtained above, we calculate that the absolute magnitudes of the progenitor of SN 2021hpr should be fainter than -4.96 mag , -5.11 mag, and -5.59 mag in the above three bands.

\section{Conclusion} \label{sec:conclusion}

In this article, we present the optical photometry and spectroscopy of SN 2021hpr, especially the early spectrum obtained at $t\sim -14.372$ days before the maximum. 
All photometric and spectroscopic data will be available on WISeREP3 \citep{2012PASP..124..668Y}. 
The derived photometric parameters of SN 2021hpr are listed in Table \ref{tab:param}.
We draw the following conclusions.

(1) The $BVRI$ light curves show that SN 2021hpr is a normal type Ia SN. Its light-curve evolution resembles that of some well-sampled SNe Ia. 
SN 2021hpr has a rise time $t_{rise}^B=16.535\pm 0.078$ days, and a post-maximum magnitude decline rate $\Delta m_{15}(B)=0.949\pm 0.019$ mag, which is lower than the typical value for normal SNe Ia. 
By about two weeks before maximum, there was no flux excess detected in the optical light curves of SN 2021hpr.

(2) The color evolution of SN 2021hpr is similar to that of some well-sampled SNe Ia during most of the time, except for a plateau phase that appeared in $V-R$ color curve before it reached the peak luminosity, which is similar to that of SN 2017cbv. 

(3) The expansion velocity of Si {\sc ii} $\lambda 6355$ of SN 2021hpr is about 12,420 km s$^{-1}$ at $t\sim -$4.31 days, which is slightly higher than that of NV SNe Ia, indicating that SN 2021hpr might be a transitional object between HV and NV SNe Ia. 
SN 2021hpr is also an HVG SN Ia, whose velocity gradient of Si {\sc ii} $\lambda 6355$ reached 800 km s$^{-1}$ day$^{-1}$ from $t\sim -14.37$ to $\sim -4.31$ days relative to the $B$-band maximum.  

(4) Including SN 2021hpr, there have been six SNe observed in NGC 3147 over the past 50 yr.
Compared with the estimation by \cite{2009ApJ...698.1307T}, the SN rate of NGC 3147 in the past 50 yr is somewhat higher than expected for SNe Ia, but lower for SNe Ibc and SNe II. 

(5) Based on the ratio of [N {\sc ii}] $\lambda$6584 and H$\alpha$ $\lambda$6563 of the star-formation regions near SN 2021hpr, the circumstellar environment of SN 2021hpr is metal-rich (12 + log(O/H) $\approx$ 8.648), comparable to the solar metallicity. 
This supports the idea that metal-rich environments are ideal sites for HV SNe. 

(6) The distance modulus of the host galaxy NGC 3147 is calculated to be 33.46 $\pm$ 0.21 mag according to the light curve of SN 2021hpr, consistent with measures for the other SNe Ia in this galaxy. 

To explore the progenitor system, it is necessary to detect a large sample of SNe Ia in the earliest phase. 
The future SiTian Project promoted by NAOC \citep{2021AnABC..93..628L} will provide us such an opportunity. 

\begin{acknowledgments}

We acknowledge the support of the staff of the Xinglong 2.16 m telescope and 60cm telescope. 
We acknowledge helpful discussions with Dr. Minyi Lin and Dr. Chaojian Wu. 
This work is partially supported by the Open Project Program of the Key Laboratory of Optical Astronomy, National Astronomical Observatories, Chinese Academy of Sciences.
This research is supported by the National Key R\&D Program of China (No. 2017YFA0402704). 
This work is supported by the National Natural Science Foundation of China (NSFC; grant Nos. 11733006, 12090041, and 12090040). 
We acknowledge the science research grants from the China Manned Space Project with NO.CMS-CSST-2021-B06 and CMS-CSST-2021-A12.

\end{acknowledgments}

\vspace{5mm}
\facilities{2.16m telescope (XingLong Observatory, NAOC), 60cm telescope (XingLong Observatory, NAOC)}

\software{Matplotlib \citep{Hunter:2007}, Numpy \citep{harris2020array}, Sicpy \citep{2020SciPy-NMeth}, Astropy \citep{2013A&A...558A..33A, 2018AJ....156..123A}, IRAF \citep{1999ascl.soft11002N}, ccdproc \citep{matt_craig_2017_1069648},  Photutils \citep{larry_bradley_2020_4044744}, Astrometry.net \citep{2012ascl.soft08001L}, reproject \citep{2020ascl.soft11023R}, SNooPy \citep{2011AJ....141...19B, 2014ApJ...789...32B}
}

\bibliography{sample631}{}

\begin{thebibliography}{}
\expandafter\ifx\csname natexlab\endcsname\relax\def\natexlab#1{#1}\fi
\providecommand{\url}[1]{\href{#1}{#1}}
\providecommand{\dodoi}[1]{doi:~\href{http://doi.org/#1}{\nolinkurl{#1}}}
\providecommand{\doeprint}[1]{\href{http://ascl.net/#1}{\nolinkurl{http://ascl.net/#1}}}
\providecommand{\doarXiv}[1]{\href{https://arxiv.org/abs/#1}{\nolinkurl{https://arxiv.org/abs/#1}}}

\bibitem[{{Amanullah} {et~al.}(2010){Amanullah}, {Lidman}, {Rubin}, {Aldering},
  {Astier}, {Barbary}, {Burns}, {Conley}, {Dawson}, {Deustua}, {Doi}, {Fabbro},
  {Faccioli}, {Fakhouri}, {Folatelli}, {Fruchter}, {Furusawa}, {Garavini},
  {Goldhaber}, {Goobar}, {Groom}, {Hook}, {Howell}, {Kashikawa}, {Kim}, {Knop},
  {Kowalski}, {Linder}, {Meyers}, {Morokuma}, {Nobili}, {Nordin}, {Nugent},
  {{\"O}stman}, {Pain}, {Panagia}, {Perlmutter}, {Raux}, {Ruiz-Lapuente},
  {Spadafora}, {Strovink}, {Suzuki}, {Wang}, {Wood-Vasey}, {Yasuda}, \&
  {Supernova Cosmology Project}}]{2010ApJ...716..712A}
{Amanullah}, R., {Lidman}, C., {Rubin}, D., {et~al.} 2010, \apj, 716, 712,
  \dodoi{10.1088/0004-637X/716/1/712}

\bibitem[{{Anupama} {et~al.}(2005){Anupama}, {Sahu}, \&
  {Jose}}]{2005A&A...429..667A}
{Anupama}, G.~C., {Sahu}, D.~K., \& {Jose}, J. 2005, \aap, 429, 667,
  \dodoi{10.1051/0004-6361:20041687}

\bibitem[{{Arnett}(1969)}]{1969Ap&SS...5..180A}
{Arnett}, W.~D. 1969, \apss, 5, 180, \dodoi{10.1007/BF00650291}

\bibitem[{{Asplund} {et~al.}(2021){Asplund}, {Amarsi}, \&
  {Grevesse}}]{2021A&A...653A.141A}
{Asplund}, M., {Amarsi}, A.~M., \& {Grevesse}, N. 2021, \aap, 653, A141,
  \dodoi{10.1051/0004-6361/202140445}

\bibitem[{{Astropy Collaboration} {et~al.}(2013){Astropy Collaboration},
  {Robitaille}, {Tollerud}, {Greenfield}, {Droettboom}, {Bray}, {Aldcroft},
  {Davis}, {Ginsburg}, {Price-Whelan}, {Kerzendorf}, {Conley}, {Crighton},
  {Barbary}, {Muna}, {Ferguson}, {Grollier}, {Parikh}, {Nair}, {Unther},
  {Deil}, {Woillez}, {Conseil}, {Kramer}, {Turner}, {Singer}, {Fox}, {Weaver},
  {Zabalza}, {Edwards}, {Azalee Bostroem}, {Burke}, {Casey}, {Crawford},
  {Dencheva}, {Ely}, {Jenness}, {Labrie}, {Lim}, {Pierfederici}, {Pontzen},
  {Ptak}, {Refsdal}, {Servillat}, \& {Streicher}}]{2013A&A...558A..33A}
{Astropy Collaboration}, {Robitaille}, T.~P., {Tollerud}, E.~J., {et~al.} 2013,
  \aap, 558, A33, \dodoi{10.1051/0004-6361/201322068}

\bibitem[{{Astropy Collaboration} {et~al.}(2018){Astropy Collaboration},
  {Price-Whelan}, {Sip{\H{o}}cz}, {G{\"u}nther}, {Lim}, {Crawford}, {Conseil},
  {Shupe}, {Craig}, {Dencheva}, {Ginsburg}, {VanderPlas}, {Bradley},
  {P{\'e}rez-Su{\'a}rez}, {de Val-Borro}, {Aldcroft}, {Cruz}, {Robitaille},
  {Tollerud}, {Ardelean}, {Babej}, {Bach}, {Bachetti}, {Bakanov}, {Bamford},
  {Barentsen}, {Barmby}, {Baumbach}, {Berry}, {Biscani}, {Boquien}, {Bostroem},
  {Bouma}, {Brammer}, {Bray}, {Breytenbach}, {Buddelmeijer}, {Burke},
  {Calderone}, {Cano Rodr{\'\i}guez}, {Cara}, {Cardoso}, {Cheedella}, {Copin},
  {Corrales}, {Crichton}, {D'Avella}, {Deil}, {Depagne}, {Dietrich}, {Donath},
  {Droettboom}, {Earl}, {Erben}, {Fabbro}, {Ferreira}, {Finethy}, {Fox},
  {Garrison}, {Gibbons}, {Goldstein}, {Gommers}, {Greco}, {Greenfield},
  {Groener}, {Grollier}, {Hagen}, {Hirst}, {Homeier}, {Horton}, {Hosseinzadeh},
  {Hu}, {Hunkeler}, {Ivezi{\'c}}, {Jain}, {Jenness}, {Kanarek}, {Kendrew},
  {Kern}, {Kerzendorf}, {Khvalko}, {King}, {Kirkby}, {Kulkarni}, {Kumar},
  {Lee}, {Lenz}, {Littlefair}, {Ma}, {Macleod}, {Mastropietro}, {McCully},
  {Montagnac}, {Morris}, {Mueller}, {Mumford}, {Muna}, {Murphy}, {Nelson},
  {Nguyen}, {Ninan}, {N{\"o}the}, {Ogaz}, {Oh}, {Parejko}, {Parley}, {Pascual},
  {Patil}, {Patil}, {Plunkett}, {Prochaska}, {Rastogi}, {Reddy Janga},
  {Sabater}, {Sakurikar}, {Seifert}, {Sherbert}, {Sherwood-Taylor}, {Shih},
  {Sick}, {Silbiger}, {Singanamalla}, {Singer}, {Sladen}, {Sooley},
  {Sornarajah}, {Streicher}, {Teuben}, {Thomas}, {Tremblay}, {Turner},
  {Terr{\'o}n}, {van Kerkwijk}, {de la Vega}, {Watkins}, {Weaver}, {Whitmore},
  {Woillez}, {Zabalza}, \& {Astropy Contributors}}]{2018AJ....156..123A}
{Astropy Collaboration}, {Price-Whelan}, A.~M., {Sip{\H{o}}cz}, B.~M., {et~al.}
  2018, \aj, 156, 123, \dodoi{10.3847/1538-3881/aabc4f}

\bibitem[{{Benetti} {et~al.}(2005){Benetti}, {Cappellaro}, {Mazzali},
  {Turatto}, {Altavilla}, {Bufano}, {Elias-Rosa}, {Kotak}, {Pignata}, {Salvo},
  \& {Stanishev}}]{2005ApJ...623.1011B}
{Benetti}, S., {Cappellaro}, E., {Mazzali}, P.~A., {et~al.} 2005, \apj, 623,
  1011, \dodoi{10.1086/428608}

\bibitem[{{Bianchi} {et~al.}(2019){Bianchi}, {Antonucci}, {Capetti},
  {Chiaberge}, {Laor}, {Bassani}, {Carrera}, {La Franca}, {Marinucci}, {Matt},
  {Middei}, \& {Panessa}}]{2019MNRAS.488L...1B}
{Bianchi}, S., {Antonucci}, R., {Capetti}, A., {et~al.} 2019, \mnras, 488, L1,
  \dodoi{10.1093/mnrasl/slz080}

\bibitem[{{Biscardi} {et~al.}(2012){Biscardi}, {Brocato}, {Arkharov}, {Di
  Carlo}, {di Rico}, {Dolci}, {Efimova}, {Pietrinferni}, \&
  {Valentini}}]{2012A&A...537A..57B}
{Biscardi}, I., {Brocato}, E., {Arkharov}, A., {et~al.} 2012, \aap, 537, A57,
  \dodoi{10.1051/0004-6361/201014160}

\bibitem[{{Blanc} {et~al.}(2013){Blanc}, {Weinzirl}, {Song}, {Heiderman},
  {Gebhardt}, {Jogee}, {Evans}, {van den Bosch}, {Luo}, {Drory}, {Fabricius},
  {Fisher}, {Hao}, {Kaplan}, {Marinova}, {Vutisalchavakul}, \&
  {Yoachim}}]{2013AJ....145..138B}
{Blanc}, G.~A., {Weinzirl}, T., {Song}, M., {et~al.} 2013, \aj, 145, 138,
  \dodoi{10.1088/0004-6256/145/5/138}

\bibitem[{{Bochenek} {et~al.}(2018){Bochenek}, {Dwarkadas}, {Silverman}, {Fox},
  {Chevalier}, {Smith}, \& {Filippenko}}]{2018MNRAS.473..336B}
{Bochenek}, C.~D., {Dwarkadas}, V.~V., {Silverman}, J.~M., {et~al.} 2018,
  \mnras, 473, 336, \dodoi{10.1093/mnras/stx2029}

\bibitem[{Bradley {et~al.}(2020)Bradley, Sip{\H o}cz, Robitaille, Tollerud,
  Vin{\'{\i}}cius, Deil, Barbary, Wilson, Busko, G{\"u}nther, Cara, Conseil,
  Bostroem, Droettboom, Bray, Bratholm, Lim, Barentsen, Craig, Pascual, Perren,
  Greco, Donath, de~Val-Borro, Kerzendorf, Bach, Weaver, D'Eugenio, Souchereau,
  \& Ferreira}]{larry_bradley_2020_4044744}
Bradley, L., Sip{\H o}cz, B., Robitaille, T., {et~al.} 2020, astropy/photutils:
  1.0.0, 1.0.0,  Zenodo, \dodoi{10.5281/zenodo.4044744}

\bibitem[{{Branch} \& {Wheeler}(2017)}]{2017suex.book.....B}
{Branch}, D., \& {Wheeler}, J.~C. 2017, {Supernova Explosions},
  \dodoi{10.1007/978-3-662-55054-0}

\bibitem[{{Brown} {et~al.}(2014){Brown}, {Breeveld}, {Holland}, {Kuin}, \&
  {Pritchard}}]{2014Ap&SS.354...89B}
{Brown}, P.~J., {Breeveld}, A.~A., {Holland}, S., {Kuin}, P., \& {Pritchard},
  T. 2014, \apss, 354, 89, \dodoi{10.1007/s10509-014-2059-8}

\bibitem[{{Brown} {et~al.}(2012){Brown}, {Dawson}, {Harris}, {Olmstead},
  {Milne}, \& {Roming}}]{2012ApJ...749...18B}
{Brown}, P.~J., {Dawson}, K.~S., {Harris}, D.~W., {et~al.} 2012, \apj, 749, 18,
  \dodoi{10.1088/0004-637X/749/1/18}

\bibitem[{{Burke} {et~al.}(2021){Burke}, {Howell}, {Sarbadhicary}, {Sand},
  {Amaro}, {Hiramatsu}, {McCully}, {Pellegrino}, {Andrews}, {Brown}, {Itagaki},
  {Shahbandeh}, {Bostroem}, {Chomiuk}, {Hsiao}, {Smith}, \&
  {Valenti}}]{2021ApJ...919..142B}
{Burke}, J., {Howell}, D.~A., {Sarbadhicary}, S.~K., {et~al.} 2021, \apj, 919,
  142, \dodoi{10.3847/1538-4357/ac126b}

\bibitem[{{Burns} {et~al.}(2011){Burns}, {Stritzinger}, {Phillips}, {Kattner},
  {Persson}, {Madore}, {Freedman}, {Boldt}, {Campillay}, {Contreras},
  {Folatelli}, {Gonzalez}, {Krzeminski}, {Morrell}, {Salgado}, \&
  {Suntzeff}}]{2011AJ....141...19B}
{Burns}, C.~R., {Stritzinger}, M., {Phillips}, M.~M., {et~al.} 2011, \aj, 141,
  19, \dodoi{10.1088/0004-6256/141/1/19}

\bibitem[{{Burns} {et~al.}(2014){Burns}, {Stritzinger}, {Phillips}, {Hsiao},
  {Contreras}, {Persson}, {Folatelli}, {Boldt}, {Campillay}, {Castell{\'o}n},
  {Freedman}, {Madore}, {Morrell}, {Salgado}, \&
  {Suntzeff}}]{2014ApJ...789...32B}
---. 2014, \apj, 789, 32, \dodoi{10.1088/0004-637X/789/1/32}

\bibitem[{{Burns} {et~al.}(2018){Burns}, {Parent}, {Phillips}, {Stritzinger},
  {Krisciunas}, {Suntzeff}, {Hsiao}, {Contreras}, {Anais}, {Boldt}, {Busta},
  {Campillay}, {Castell{\'o}n}, {Folatelli}, {Freedman}, {Gonz{\'a}lez},
  {Hamuy}, {Heoflich}, {Krzeminski}, {Madore}, {Morrell}, {Persson}, {Roth},
  {Salgado}, {Ser{\'o}n}, \& {Torres}}]{2018ApJ...869...56B}
{Burns}, C.~R., {Parent}, E., {Phillips}, M.~M., {et~al.} 2018, \apj, 869, 56,
  \dodoi{10.3847/1538-4357/aae51c}

\bibitem[{{Cao} {et~al.}(2015){Cao}, {Kulkarni}, {Howell}, {Gal-Yam},
  {Kasliwal}, {Valenti}, {Johansson}, {Amanullah}, {Goobar}, {Sollerman},
  {Taddia}, {Horesh}, {Sagiv}, {Cenko}, {Nugent}, {Arcavi}, {Surace},
  {Wo{\'z}niak}, {Moody}, {Rebbapragada}, {Bue}, \&
  {Gehrels}}]{2015Natur.521..328C}
{Cao}, Y., {Kulkarni}, S.~R., {Howell}, D.~A., {et~al.} 2015, \nat, 521, 328,
  \dodoi{10.1038/nature14440}

\bibitem[{{Childress} {et~al.}(2014){Childress}, {Filippenko}, {Ganeshalingam},
  \& {Schmidt}}]{2014MNRAS.437..338C}
{Childress}, M.~J., {Filippenko}, A.~V., {Ganeshalingam}, M., \& {Schmidt},
  B.~P. 2014, \mnras, 437, 338, \dodoi{10.1093/mnras/stt1892}

\bibitem[{{Conley} {et~al.}(2006){Conley}, {Howell}, {Howes}, {Sullivan},
  {Astier}, {Balam}, {Basa}, {Carlberg}, {Fouchez}, {Guy}, {Hook}, {Neill},
  {Pain}, {Perrett}, {Pritchet}, {Regnault}, {Rich}, {Taillet}, {Aubourg},
  {Bronder}, {Ellis}, {Fabbro}, {Filiol}, {Le Borgne}, {Palanque-Delabrouille},
  {Perlmutter}, \& {Ripoche}}]{2006AJ....132.1707C}
{Conley}, A., {Howell}, D.~A., {Howes}, A., {et~al.} 2006, \aj, 132, 1707,
  \dodoi{10.1086/507788}

\bibitem[{{C{\^o}t{\'e}} {et~al.}(2016){C{\^o}t{\'e}}, {Ritter}, {O'Shea},
  {Herwig}, {Pignatari}, {Jones}, \& {Fryer}}]{2016ApJ...824...82C}
{C{\^o}t{\'e}}, B., {Ritter}, C., {O'Shea}, B.~W., {et~al.} 2016, \apj, 824,
  82, \dodoi{10.3847/0004-637X/824/2/82}

\bibitem[{{Cousins}(1980)}]{1980SAAOC...1..234C}
{Cousins}, A.~W.~J. 1980, South African Astronomical Observatory Circular, 1,
  234

\bibitem[{Craig {et~al.}(2017)Craig, Crawford, Seifert, Robitaille, Sip{\H
  o}cz, Walawender, Vin{\'{\i}}cius, Ninan, Droettboom, Youn, Tollerud, Bray,
  Walker, Janga, Stotts, G{\"u}nther, Rol, Bach, Bradley, Deil, Price-Whelan,
  Barbary, Horton, Schoenell, Heidt, Gasdia, Nelson, \&
  Streicher}]{matt_craig_2017_1069648}
Craig, M., Crawford, S., Seifert, M., {et~al.} 2017, astropy/ccdproc:
  v1.3.0.post1, \dodoi{10.5281/zenodo.1069648}

\bibitem[{Craig {et~al.}(2021)Craig, Crawford, Seifert, Robitaille, Sipőcz,
  Walawender, Crawford, Vinícius, Ninan, Droettboom, Youn, Yash-10, Tollerud,
  Bowers, Bray, Bach, stottsco, Janga, walkerna22, Lim, Günther, Rol, A.,
  Bradley, Price-Whelan, Deil, Ryon, Lee, Barbary, \&
  Weiner}]{matt_craig_2021_5796852}
---. 2021, astropy/ccdproc: 2.3.0 -- cosmic ray update, 2.3.0,  Zenodo,
  \dodoi{10.5281/zenodo.5796852}

\bibitem[{{De} {et~al.}(2019){De}, {Kasliwal}, {Polin}, {Nugent}, {Bildsten},
  {Adams}, {Bellm}, {Blagorodnova}, {Burdge}, {Cannella}, {Cenko}, {Dekany},
  {Feeney}, {Hale}, {Fremling}, {Graham}, {Ho}, {Jencson}, {Kulkarni}, {Laher},
  {Masci}, {Miller}, {Patterson}, {Rebbapragada}, {Riddle}, {Shupe}, \&
  {Smith}}]{2019ApJ...873L..18D}
{De}, K., {Kasliwal}, M.~M., {Polin}, A., {et~al.} 2019, \apjl, 873, L18,
  \dodoi{10.3847/2041-8213/ab0aec}

\bibitem[{{Dilday} {et~al.}(2012){Dilday}, {Howell}, {Cenko}, {Silverman},
  {Nugent}, {Sullivan}, {Ben-Ami}, {Bildsten}, {Bolte}, {Endl}, {Filippenko},
  {Gnat}, {Horesh}, {Hsiao}, {Kasliwal}, {Kirkman}, {Maguire}, {Marcy},
  {Moore}, {Pan}, {Parrent}, {Podsiadlowski}, {Quimby}, {Sternberg}, {Suzuki},
  {Tytler}, {Xu}, {Bloom}, {Gal-Yam}, {Hook}, {Kulkarni}, {Law}, {Ofek},
  {Polishook}, \& {Poznanski}}]{2012Sci...337..942D}
{Dilday}, B., {Howell}, D.~A., {Cenko}, S.~B., {et~al.} 2012, Science, 337,
  942, \dodoi{10.1126/science.1219164}

\bibitem[{{Dimitriadis} {et~al.}(2019){Dimitriadis}, {Foley}, {Rest}, {Kasen},
  {Piro}, {Polin}, {Jones}, {Villar}, {Narayan}, {Coulter}, {Kilpatrick},
  {Pan}, {Rojas-Bravo}, {Fox}, {Jha}, {Nugent}, {Riess}, {Scolnic}, {Drout},
  {K2 Mission Team}, {Barentsen}, {Dotson}, {Gully-Santiago}, {Hedges}, {Cody},
  {Barclay}, {Howell}, {KEGS}, {Garnavich}, {Tucker}, {Shaya}, {Mushotzky},
  {Olling}, {Margheim}, {Zenteno}, {Kepler spacecraft Team}, {Coughlin}, {Van
  Cleve}, {Cardoso}, {Larson}, {McCalmont-Everton}, {Peterson}, {Ross},
  {Reedy}, {Osborne}, {McGinn}, {Kohnert}, {Migliorini}, {Wheaton}, {Spencer},
  {Labonde}, {Castillo}, {Beerman}, {Steward}, {Hanley}, {Larsen},
  {Gangopadhyay}, {Kloetzel}, {Weschler}, {Nystrom}, {Moffatt}, {Redick},
  {Griest}, {Packard}, {Muszynski}, {Kampmeier}, {Bjella}, {Flynn},
  {Elsaesser}, {Pan-STARRS}, {Chambers}, {Flewelling}, {Huber}, {Magnier},
  {Waters}, {Schultz}, {Bulger}, {Lowe}, {Willman}, {Smartt}, {Smith}, {DECam},
  {Points}, {Strampelli}, {ASAS-SN}, {Brimacombe}, {Chen}, {Mu{\~n}oz},
  {Mutel}, {Shields}, {Vallely}, {Villanueva}, {PTSS/TNTS}, {Li}, {Wang},
  {Zhang}, {Lin}, {Mo}, {Zhao}, {Sai}, {Zhang}, {Zhang}, {Zhang}, {Wang},
  {Zhang}, {Baron}, {DerKacy}, {Li}, {Chen}, {Xiang}, {Rui}, {Wang}, {Huang},
  {Li}, {Cumbres Observatory}, {Hosseinzadeh}, {Howell}, {Arcavi}, {Hiramatsu},
  {Burke}, {Valenti}, {ATLAS}, {Tonry}, {Denneau}, {Heinze}, {Weiland},
  {Stalder}, {Konkoly}, {Vink{\'o}}, {S{\'a}rneczky}, {P{\'a}l}, {B{\'o}di},
  {Bogn{\'a}r}, {Cs{\'a}k}, {Cseh}, {Cs{\"o}rnyei}, {Hanyecz}, {Ign{\'a}cz},
  {Kalup}, {K{\"o}nyves-T{\'o}th}, {Kriskovics}, {Ordasi}, {Rajmon},
  {S{\'o}dor}, {Szab{\'o}}, {Szak{\'a}ts}, {Zsidi}, {ePESSTO}, {Williams},
  {Nordin}, {Cartier}, {Frohmaier}, {Galbany}, {Guti{\'e}rrez}, {Hook},
  {Inserra}, {Smith}, {Arizona}, {Sand}, {Andrews}, {Smith}, \&
  {Bilinski}}]{2019ApJ...870L...1D}
{Dimitriadis}, G., {Foley}, R.~J., {Rest}, A., {et~al.} 2019, \apjl, 870, L1,
  \dodoi{10.3847/2041-8213/aaedb0}

\bibitem[{{Duszanowicz}(2006)}]{2006IAUC.8755....3D}
{Duszanowicz}, G. 2006, \iaucirc, 8755, 3

\bibitem[{{Dwek}(1998)}]{1998ApJ...501..643D}
{Dwek}, E. 1998, \apj, 501, 643, \dodoi{10.1086/305829}

\bibitem[{{Fan} {et~al.}(2016){Fan}, {Wang}, {Jiang}, {Wu}, {Li}, {Huang},
  {Xu}, {Hu}, {Zhu}, {Wang}, {Komossa}, \& {Zhang}}]{2016PASP..128k5005F}
{Fan}, Z., {Wang}, H., {Jiang}, X., {et~al.} 2016, \pasp, 128, 115005,
  \dodoi{10.1088/1538-3873/128/969/115005}

\bibitem[{{Firth} {et~al.}(2015){Firth}, {Sullivan}, {Gal-Yam}, {Howell},
  {Maguire}, {Nugent}, {Piro}, {Baltay}, {Feindt}, {Hadjiyksta}, {McKinnon},
  {Ofek}, {Rabinowitz}, \& {Walker}}]{2015MNRAS.446.3895F}
{Firth}, R.~E., {Sullivan}, M., {Gal-Yam}, A., {et~al.} 2015, \mnras, 446,
  3895, \dodoi{10.1093/mnras/stu2314}

\bibitem[{{Foley}(2012)}]{2012ApJ...748..127F}
{Foley}, R.~J. 2012, \apj, 748, 127, \dodoi{10.1088/0004-637X/748/2/127}

\bibitem[{{Foley} \& {Kasen}(2011)}]{2011ApJ...729...55F}
{Foley}, R.~J., \& {Kasen}, D. 2011, \apj, 729, 55,
  \dodoi{10.1088/0004-637X/729/1/55}

\bibitem[{{Foley} {et~al.}(2011){Foley}, {Sanders}, \&
  {Kirshner}}]{2011ApJ...742...89F}
{Foley}, R.~J., {Sanders}, N.~E., \& {Kirshner}, R.~P. 2011, \apj, 742, 89,
  \dodoi{10.1088/0004-637X/742/2/89}

\bibitem[{{Foley} {et~al.}(2012){Foley}, {Challis}, {Filippenko},
  {Ganeshalingam}, {Landsman}, {Li}, {Marion}, {Silverman}, {Beaton},
  {Bennert}, {Cenko}, {Childress}, {Guhathakurta}, {Jiang}, {Kalirai},
  {Kirshner}, {Stockton}, {Tollerud}, {Vink{\'o}}, {Wheeler}, \&
  {Woo}}]{2012ApJ...744...38F}
{Foley}, R.~J., {Challis}, P.~J., {Filippenko}, A.~V., {et~al.} 2012, \apj,
  744, 38, \dodoi{10.1088/0004-637X/744/1/38}

\bibitem[{{F{\"o}rster} {et~al.}(2021){F{\"o}rster}, {Cabrera-Vives},
  {Castillo-Navarrete}, {Est{\'e}vez}, {S{\'a}nchez-S{\'a}ez}, {Arredondo},
  {Bauer}, {Carrasco-Davis}, {Catelan}, {Elorrieta}, {Eyheramendy}, {Huijse},
  {Pignata}, {Reyes}, {Reyes}, {Rodr{\'\i}guez-Mancini}, {Ruz-Mieres},
  {Valenzuela}, {{\'A}lvarez-Maldonado}, {Astorga}, {Borissova}, {Clocchiatti},
  {De Cicco}, {Donoso-Oliva}, {Hern{\'a}ndez-Garc{\'\i}a}, {Graham},
  {Jord{\'a}n}, {Kurtev}, {Mahabal}, {Maureira}, {Mu{\~n}oz-Arancibia},
  {Molina-Ferreiro}, {Moya}, {Palma}, {P{\'e}rez-Carrasco}, {Protopapas},
  {Romero}, {Sabatini-Gacitua}, {S{\'a}nchez}, {San Mart{\'\i}n},
  {Sep{\'u}lveda-Cobo}, {Vera}, \& {Vergara}}]{2021AJ....161..242F}
{F{\"o}rster}, F., {Cabrera-Vives}, G., {Castillo-Navarrete}, E., {et~al.}
  2021, \aj, 161, 242, \dodoi{10.3847/1538-3881/abe9bc}

\bibitem[{{Freedman} {et~al.}(2019){Freedman}, {Madore}, {Hatt}, {Hoyt},
  {Jang}, {Beaton}, {Burns}, {Lee}, {Monson}, {Neeley}, {Phillips}, {Rich}, \&
  {Seibert}}]{2019ApJ...882...34F}
{Freedman}, W.~L., {Madore}, B.~F., {Hatt}, D., {et~al.} 2019, \apj, 882, 34,
  \dodoi{10.3847/1538-4357/ab2f73}

\bibitem[{{Ganeshalingam} {et~al.}(2011){Ganeshalingam}, {Li}, \&
  {Filippenko}}]{2011MNRAS.416.2607G}
{Ganeshalingam}, M., {Li}, W., \& {Filippenko}, A.~V. 2011, \mnras, 416, 2607,
  \dodoi{10.1111/j.1365-2966.2011.19213.x}

\bibitem[{{Ganeshalingam} {et~al.}(2010){Ganeshalingam}, {Li}, {Filippenko},
  {Anderson}, {Foster}, {Gates}, {Griffith}, {Grigsby}, {Joubert}, {Leja},
  {Lowe}, {Macomber}, {Pritchard}, {Thrasher}, \&
  {Winslow}}]{2010ApJS..190..418G}
{Ganeshalingam}, M., {Li}, W., {Filippenko}, A.~V., {et~al.} 2010, \apjs, 190,
  418, \dodoi{10.1088/0067-0049/190/2/418}

\bibitem[{{Garg} {et~al.}(2007){Garg}, {Stubbs}, {Challis}, {Wood-Vasey},
  {Blondin}, {Huber}, {Cook}, {Nikolaev}, {Rest}, {Smith}, {Olsen}, {Suntzeff},
  {Aguilera}, {Prieto}, {Becker}, {Miceli}, {Miknaitis}, {Clocchiatti},
  {Minniti}, {Morelli}, \& {Welch}}]{2007AJ....133..403G}
{Garg}, A., {Stubbs}, C.~W., {Challis}, P., {et~al.} 2007, \aj, 133, 403,
  \dodoi{10.1086/510118}

\bibitem[{{Gil de Paz} {et~al.}(2007){Gil de Paz}, {Boissier}, {Madore},
  {Seibert}, {Joe}, {Boselli}, {Wyder}, {Thilker}, {Bianchi}, {Rey}, {Rich},
  {Barlow}, {Conrow}, {Forster}, {Friedman}, {Martin}, {Morrissey}, {Neff},
  {Schiminovich}, {Small}, {Donas}, {Heckman}, {Lee}, {Milliard}, {Szalay}, \&
  {Yi}}]{2007ApJS..173..185G}
{Gil de Paz}, A., {Boissier}, S., {Madore}, B.~F., {et~al.} 2007, \apjs, 173,
  185, \dodoi{10.1086/516636}

\bibitem[{{Goldhaber} {et~al.}(2001){Goldhaber}, {Groom}, {Kim}, {Aldering},
  {Astier}, {Conley}, {Deustua}, {Ellis}, {Fabbro}, {Fruchter}, {Goobar},
  {Hook}, {Irwin}, {Kim}, {Knop}, {Lidman}, {McMahon}, {Nugent}, {Pain},
  {Panagia}, {Pennypacker}, {Perlmutter}, {Ruiz-Lapuente}, {Schaefer},
  {Walton}, \& {York}}]{2001ApJ...558..359G}
{Goldhaber}, G., {Groom}, D.~E., {Kim}, A., {et~al.} 2001, \apj, 558, 359,
  \dodoi{10.1086/322460}

\bibitem[{{Gronow} {et~al.}(2020){Gronow}, {Collins}, {Ohlmann}, {Pakmor},
  {Kromer}, {Seitenzahl}, {Sim}, \& {R{\"o}pke}}]{2020A&A...635A.169G}
{Gronow}, S., {Collins}, C., {Ohlmann}, S.~T., {et~al.} 2020, \aap, 635, A169,
  \dodoi{10.1051/0004-6361/201936494}

\bibitem[{{Guy} {et~al.}(2005){Guy}, {Astier}, {Nobili}, {Regnault}, \&
  {Pain}}]{2005A&A...443..781G}
{Guy}, J., {Astier}, P., {Nobili}, S., {Regnault}, N., \& {Pain}, R. 2005,
  \aap, 443, 781, \dodoi{10.1051/0004-6361:20053025}

\bibitem[{Harris {et~al.}(2020)Harris, Millman, van~der Walt, Gommers,
  Virtanen, Cournapeau, Wieser, Taylor, Berg, Smith, Kern, Picus, Hoyer, van
  Kerkwijk, Brett, Haldane, del R{\'{i}}o, Wiebe, Peterson,
  G{\'{e}}rard-Marchant, Sheppard, Reddy, Weckesser, Abbasi, Gohlke, \&
  Oliphant}]{harris2020array}
Harris, C.~R., Millman, K.~J., van~der Walt, S.~J., {et~al.} 2020, Nature, 585,
  357, \dodoi{10.1038/s41586-020-2649-2}

\bibitem[{{Hayden} {et~al.}(2010){Hayden}, {Garnavich}, {Kessler}, {Frieman},
  {Jha}, {Bassett}, {Cinabro}, {Dilday}, {Kasen}, {Marriner}, {Nichol},
  {Riess}, {Sako}, {Schneider}, {Smith}, \& {Sollerman}}]{2010ApJ...712..350H}
{Hayden}, B.~T., {Garnavich}, P.~M., {Kessler}, R., {et~al.} 2010, \apj, 712,
  350, \dodoi{10.1088/0004-637X/712/1/350}

\bibitem[{{Hicken} {et~al.}(2009){Hicken}, {Challis}, {Jha}, {Kirshner},
  {Matheson}, {Modjaz}, {Rest}, {Wood-Vasey}, {Bakos}, {Barton}, {Berlind},
  {Bragg}, {Brice{\~n}o}, {Brown}, {Caldwell}, {Calkins}, {Cho}, {Ciupik},
  {Contreras}, {Dendy}, {Dosaj}, {Durham}, {Eriksen}, {Esquerdo}, {Everett},
  {Falco}, {Fernandez}, {Gaba}, {Garnavich}, {Graves}, {Green}, {Groner},
  {Hergenrother}, {Holman}, {Hradecky}, {Huchra}, {Hutchison}, {Jerius},
  {Jordan}, {Kilgard}, {Krauss}, {Luhman}, {Macri}, {Marrone}, {McDowell},
  {McIntosh}, {McNamara}, {Megeath}, {Mochejska}, {Munoz}, {Muzerolle},
  {Naranjo}, {Narayan}, {Pahre}, {Peters}, {Peterson}, {Rines}, {Ripman},
  {Roussanova}, {Schild}, {Sicilia-Aguilar}, {Sokoloski}, {Smalley}, {Smith},
  {Spahr}, {Stanek}, {Barmby}, {Blondin}, {Stubbs}, {Szentgyorgyi}, {Torres},
  {Vaz}, {Vikhlinin}, {Wang}, {Westover}, {Woods}, \&
  {Zhao}}]{2009ApJ...700..331H}
{Hicken}, M., {Challis}, P., {Jha}, S., {et~al.} 2009, \apj, 700, 331,
  \dodoi{10.1088/0004-637X/700/1/331}

\bibitem[{{Hicken} {et~al.}(2012){Hicken}, {Challis}, {Kirshner}, {Rest},
  {Cramer}, {Wood-Vasey}, {Bakos}, {Berlind}, {Brown}, {Caldwell}, {Calkins},
  {Currie}, {de Kleer}, {Esquerdo}, {Everett}, {Falco}, {Fernandez},
  {Friedman}, {Groner}, {Hartman}, {Holman}, {Hutchins}, {Keys}, {Kipping},
  {Latham}, {Marion}, {Narayan}, {Pahre}, {Pal}, {Peters}, {Perumpilly},
  {Ripman}, {Sipocz}, {Szentgyorgyi}, {Tang}, {Torres}, {Vaz}, {Wolk}, \&
  {Zezas}}]{2012ApJS..200...12H}
{Hicken}, M., {Challis}, P., {Kirshner}, R.~P., {et~al.} 2012, \apjs, 200, 12,
  \dodoi{10.1088/0067-0049/200/2/12}

\bibitem[{{Hosseinzadeh} {et~al.}(2017){Hosseinzadeh}, {Sand}, {Valenti},
  {Brown}, {Howell}, {McCully}, {Kasen}, {Arcavi}, {Bostroem}, {Tartaglia},
  {Hsiao}, {Davis}, {Shahbandeh}, \& {Stritzinger}}]{2017ApJ...845L..11H}
{Hosseinzadeh}, G., {Sand}, D.~J., {Valenti}, S., {et~al.} 2017, \apjl, 845,
  L11, \dodoi{10.3847/2041-8213/aa8402}

\bibitem[{Hunter(2007)}]{Hunter:2007}
Hunter, J.~D. 2007, Computing in Science \& Engineering, 9, 90,
  \dodoi{10.1109/MCSE.2007.55}

\bibitem[{{Iben} \& {Tutukov}(1984)}]{1984ApJS...54..335I}
{Iben}, I., J., \& {Tutukov}, A.~V. 1984, \apjs, 54, 335,
  \dodoi{10.1086/190932}

\bibitem[{{Itagaki}(2021)}]{2021TNSTR.998....1I}
{Itagaki}, K. 2021, Transient Name Server Discovery Report, 2021-998, 1

\bibitem[{{Jha} {et~al.}(2007){Jha}, {Riess}, \&
  {Kirshner}}]{2007ApJ...659..122J}
{Jha}, S., {Riess}, A.~G., \& {Kirshner}, R.~P. 2007, \apj, 659, 122,
  \dodoi{10.1086/512054}

\bibitem[{{Jha} {et~al.}(2006){Jha}, {Kirshner}, {Challis}, {Garnavich},
  {Matheson}, {Soderberg}, {Graves}, {Hicken}, {Alves}, {Arce}, {Balog},
  {Barmby}, {Barton}, {Berlind}, {Bragg}, {Brice{\~n}o}, {Brown}, {Buckley},
  {Caldwell}, {Calkins}, {Carter}, {Concannon}, {Donnelly}, {Eriksen},
  {Fabricant}, {Falco}, {Fiore}, {Garcia}, {G{\'o}mez}, {Grogin}, {Groner},
  {Groot}, {Haisch}, {Hartmann}, {Hergenrother}, {Holman}, {Huchra},
  {Jayawardhana}, {Jerius}, {Kannappan}, {Kim}, {Kleyna}, {Kochanek},
  {Koranyi}, {Krockenberger}, {Lada}, {Luhman}, {Luu}, {Macri}, {Mader},
  {Mahdavi}, {Marengo}, {Marsden}, {McLeod}, {McNamara}, {Megeath}, {Moraru},
  {Mossman}, {Muench}, {Mu{\~n}oz}, {Muzerolle}, {Naranjo}, {Nelson-Patel},
  {Pahre}, {Patten}, {Peters}, {Peters}, {Raymond}, {Rines}, {Schild},
  {Sobczak}, {Spahr}, {Stauffer}, {Stefanik}, {Szentgyorgyi}, {Tollestrup},
  {V{\"a}is{\"a}nen}, {Vikhlinin}, {Wang}, {Willner}, {Wolk}, {Zajac}, {Zhao},
  \& {Stanek}}]{2006AJ....131..527J}
{Jha}, S., {Kirshner}, R.~P., {Challis}, P., {et~al.} 2006, \aj, 131, 527,
  \dodoi{10.1086/497989}

\bibitem[{{Jiang} {et~al.}(2017){Jiang}, {Doi}, {Maeda}, {Shigeyama}, {Nomoto},
  {Yasuda}, {Jha}, {Tanaka}, {Morokuma}, {Tominaga}, {Ivezi{\'c}},
  {Ruiz-Lapuente}, {Stritzinger}, {Mazzali}, {Ashall}, {Mould}, {Baade},
  {Suzuki}, {Connolly}, {Patat}, {Wang}, {Yoachim}, {Jones}, {Furusawa}, \&
  {Miyazaki}}]{2017Natur.550...80J}
{Jiang}, J.-A., {Doi}, M., {Maeda}, K., {et~al.} 2017, \nat, 550, 80,
  \dodoi{10.1038/nature23908}

\bibitem[{{Johnson} {et~al.}(1966){Johnson}, {Mitchell}, {Iriarte}, \&
  {Wisniewski}}]{1966CoLPL...4...99J}
{Johnson}, H.~L., {Mitchell}, R.~I., {Iriarte}, B., \& {Wisniewski}, W.~Z.
  1966, Communications of the Lunar and Planetary Laboratory, 4, 99

\bibitem[{{Kasen}(2010)}]{2010ApJ...708.1025K}
{Kasen}, D. 2010, \apj, 708, 1025, \dodoi{10.1088/0004-637X/708/2/1025}

\bibitem[{{Khokhlov}(1991)}]{1991A&A...245..114K}
{Khokhlov}, A.~M. 1991, \aap, 245, 114

\bibitem[{{Kobayashi} \& {Nomoto}(2009)}]{2009ApJ...707.1466K}
{Kobayashi}, C., \& {Nomoto}, K. 2009, \apj, 707, 1466,
  \dodoi{10.1088/0004-637X/707/2/1466}

\bibitem[{{Kobayashi} {et~al.}(1998){Kobayashi}, {Tsujimoto}, {Nomoto},
  {Hachisu}, \& {Kato}}]{1998ApJ...503L.155K}
{Kobayashi}, C., {Tsujimoto}, T., {Nomoto}, K., {Hachisu}, I., \& {Kato}, M.
  1998, \apjl, 503, L155, \dodoi{10.1086/311556}

\bibitem[{{Kowalski} {et~al.}(2008){Kowalski}, {Rubin}, {Aldering},
  {Agostinho}, {Amadon}, {Amanullah}, {Balland}, {Barbary}, {Blanc}, {Challis},
  {Conley}, {Connolly}, {Covarrubias}, {Dawson}, {Deustua}, {Ellis}, {Fabbro},
  {Fadeyev}, {Fan}, {Farris}, {Folatelli}, {Frye}, {Garavini}, {Gates},
  {Germany}, {Goldhaber}, {Goldman}, {Goobar}, {Groom}, {Haissinski}, {Hardin},
  {Hook}, {Kent}, {Kim}, {Knop}, {Lidman}, {Linder}, {Mendez}, {Meyers},
  {Miller}, {Moniez}, {Mour{\~a}o}, {Newberg}, {Nobili}, {Nugent}, {Pain},
  {Perdereau}, {Perlmutter}, {Phillips}, {Prasad}, {Quimby}, {Regnault},
  {Rich}, {Rubenstein}, {Ruiz-Lapuente}, {Santos}, {Schaefer}, {Schommer},
  {Smith}, {Soderberg}, {Spadafora}, {Strolger}, {Strovink}, {Suntzeff},
  {Suzuki}, {Thomas}, {Walton}, {Wang}, {Wood-Vasey}, \&
  {Yun}}]{2008ApJ...686..749K}
{Kowalski}, M., {Rubin}, D., {Aldering}, G., {et~al.} 2008, \apj, 686, 749,
  \dodoi{10.1086/589937}

\bibitem[{{Krisciunas} {et~al.}(2003){Krisciunas}, {Suntzeff}, {Candia},
  {Arenas}, {Espinoza}, {Gonzalez}, {Gonzalez}, {H{\"o}flich}, {Landolt},
  {Phillips}, \& {Pizarro}}]{2003AJ....125..166K}
{Krisciunas}, K., {Suntzeff}, N.~B., {Candia}, P., {et~al.} 2003, \aj, 125,
  166, \dodoi{10.1086/345571}

\bibitem[{{Kromer} {et~al.}(2016){Kromer}, {Fremling}, {Pakmor},
  {Taubenberger}, {Amanullah}, {Cenko}, {Fransson}, {Goobar}, {Leloudas},
  {Taddia}, {R{\"o}pke}, {Seitenzahl}, {Sim}, \&
  {Sollerman}}]{2016MNRAS.459.4428K}
{Kromer}, M., {Fremling}, C., {Pakmor}, R., {et~al.} 2016, \mnras, 459, 4428,
  \dodoi{10.1093/mnras/stw962}

\bibitem[{{Lacchin} {et~al.}(2021){Lacchin}, {Calura}, \&
  {Vesperini}}]{2021MNRAS.506.5951L}
{Lacchin}, E., {Calura}, F., \& {Vesperini}, E. 2021, \mnras, 506, 5951,
  \dodoi{10.1093/mnras/stab2061}

\bibitem[{{Lang} {et~al.}(2012){Lang}, {Hogg}, {Mierle}, {Blanton}, \&
  {Roweis}}]{2012ascl.soft08001L}
{Lang}, D., {Hogg}, D.~W., {Mierle}, K., {Blanton}, M., \& {Roweis}, S. 2012,
  {Astrometry.net: Astrometric calibration of images}.
\newblock \doeprint{1208.001}

\bibitem[{{Leibundgut} \& {Pinto}(1992)}]{1992ApJ...401...49L}
{Leibundgut}, B., \& {Pinto}, P.~A. 1992, \apj, 401, 49, \dodoi{10.1086/172037}

\bibitem[{{Li} {et~al.}(2011){Li}, {Bloom}, {Podsiadlowski}, {Miller}, {Cenko},
  {Jha}, {Sullivan}, {Howell}, {Nugent}, {Butler}, {Ofek}, {Kasliwal},
  {Richards}, {Stockton}, {Shih}, {Bildsten}, {Shara}, {Bibby}, {Filippenko},
  {Ganeshalingam}, {Silverman}, {Kulkarni}, {Law}, {Poznanski}, {Quimby},
  {McCully}, {Patel}, {Maguire}, \& {Shen}}]{2011Natur.480..348L}
{Li}, W., {Bloom}, J.~S., {Podsiadlowski}, P., {et~al.} 2011, \nat, 480, 348,
  \dodoi{10.1038/nature10646}

\bibitem[{{Li} {et~al.}(2019){Li}, {Wang}, {Vink{\'o}}, {Mo}, {Hosseinzadeh},
  {Sand}, {Zhang}, {Lin}, {PTSS/TNTS}, {Zhang}, {Wang}, {Zhang}, {Chen},
  {Xiang}, {Rui}, {Huang}, {Li}, {Zhang}, {Li}, {Baron}, {Derkacy}, {Zhao},
  {Sai}, {Zhang}, {Wang}, {LCO}, {Howell}, {McCully}, {Arcavi}, {Valenti},
  {Hiramatsu}, {Burke}, {KEGS}, {Rest}, {Garnavich}, {Tucker}, {Narayan},
  {Shaya}, {Margheim}, {Zenteno}, {Villar}, {UCSC}, {Dimitriadis}, {Foley},
  {Pan}, {Coulter}, {Fox}, {Jha}, {Jones}, {Kasen}, {Kilpatrick}, {Piro},
  {Riess}, {Rojas-Bravo}, {ASAS-SN}, {Shappee}, {Holoien}, {Stanek}, {Drout},
  {Auchettl}, {Kochanek}, {Brown}, {Bose}, {Bersier}, {Brimacombe}, {Chen},
  {Dong}, {Holmbo}, {Mu{\~n}oz}, {Mutel}, {Post}, {Prieto}, {Shields},
  {Tallon}, {Thompson}, {Vallely}, {Villanueva}, {Pan-STARRS}, {Smartt},
  {Smith}, {Chambers}, {Flewelling}, {Huber}, {Magnier}, {Waters}, {Schultz},
  {Bulger}, {Lowe}, {Willman}, {Konkoly/Texas}, {S{\'a}rneczky}, {P{\'a}l},
  {Wheeler}, {B{\'o}di}, {Bogn{\'a}r}, {Cs{\'a}k}, {Cseh}, {Cs{\"o}rnyei},
  {Hanyecz}, {Ign{\'a}cz}, {Kalup}, {K{\"o}nyves-T{\'o}th}, {Kriskovics},
  {Ordasi}, {Rajmon}, {S{\'o}dor}, {Szab{\'o}}, {Szak{\'a}ts}, {Zsidi},
  {Arizona}, {Milne}, {Andrews}, {Smith}, {Bilinski}, {Swift}, {Brown},
  {ePESSTO}, {Nordin}, {Williams}, {Galbany}, {Palmerio}, {Hook}, {Inserra},
  {Maguire}, {Cartier}, {Razza}, {Guti{\'e}rrez}, {North Carolina}, {Hermes},
  {Reding}, {Kaiser}, {ATLAS}, {Tonry}, {Heinze}, {Denneau}, {Weiland},
  {Stalder}, {K2 Mission Team}, {Barentsen}, {Dotson}, {Barclay},
  {Gully-Santiago}, {Hedges}, {Cody}, {Howell}, {Kepler Spacecraft Team},
  {Coughlin}, {Van Cleve}, {Cardoso}, {Larson}, {McCalmont-Everton},
  {Peterson}, {Ross}, {Reedy}, {Osborne}, {McGinn}, {Kohnert}, {Migliorini},
  {Wheaton}, {Spencer}, {Labonde}, {Castillo}, {Beerman}, {Steward}, {Hanley},
  {Larsen}, {Gangopadhyay}, {Kloetzel}, {Weschler}, {Nystrom}, {Moffatt},
  {Redick}, {Griest}, {Packard}, {Muszynski}, {Kampmeier}, {Bjella}, {Flynn},
  \& {Elsaesser}}]{2019ApJ...870...12L}
{Li}, W., {Wang}, X., {Vink{\'o}}, J., {et~al.} 2019, \apj, 870, 12,
  \dodoi{10.3847/1538-4357/aaec74}

\bibitem[{{Li} {et~al.}(2021){Li}, {Wang}, {Bulla}, {Pan}, {Wang}, {Mo},
  {Zhang}, {Wu}, {Zhang}, {Zhang}, {Xiang}, {Lin}, {Sai}, {Zhang}, {Chen}, \&
  {Yan}}]{2021ApJ...906...99L}
{Li}, W., {Wang}, X., {Bulla}, M., {et~al.} 2021, \apj, 906, 99,
  \dodoi{10.3847/1538-4357/abc9b5}

\bibitem[{{Liu} {et~al.}(2021){Liu}, {Soria}, {Wu}, {Wu}, \&
  {Shang}}]{2021AnABC..93..628L}
{Liu}, J., {Soria}, R., {Wu}, X.-F., {Wu}, H., \& {Shang}, Z. 2021, An. Acad.
  Bras. Ci{\^e}nc. vol.93 supl.1, 93, 20200628,
  \dodoi{10.1590/0001-3765202120200628}

\bibitem[{{Lundqvist} {et~al.}(2015){Lundqvist}, {Nyholm}, {Taddia},
  {Sollerman}, {Johansson}, {Kozma}, {Lundqvist}, {Fransson}, {Garnavich},
  {Kromer}, {Shappee}, \& {Goobar}}]{2015A&A...577A..39L}
{Lundqvist}, P., {Nyholm}, A., {Taddia}, F., {et~al.} 2015, \aap, 577, A39,
  \dodoi{10.1051/0004-6361/201525719}

\bibitem[{{Maeda} {et~al.}(2010){Maeda}, {Benetti}, {Stritzinger}, {R{\"o}pke},
  {Folatelli}, {Sollerman}, {Taubenberger}, {Nomoto}, {Leloudas}, {Hamuy},
  {Tanaka}, {Mazzali}, \& {Elias-Rosa}}]{2010Natur.466...82M}
{Maeda}, K., {Benetti}, S., {Stritzinger}, M., {et~al.} 2010, \nat, 466, 82,
  \dodoi{10.1038/nature09122}

\bibitem[{{Maeda} {et~al.}(2011){Maeda}, {Leloudas}, {Taubenberger},
  {Stritzinger}, {Sollerman}, {Elias-Rosa}, {Benetti}, {Hamuy}, {Folatelli}, \&
  {Mazzali}}]{2011MNRAS.413.3075M}
{Maeda}, K., {Leloudas}, G., {Taubenberger}, S., {et~al.} 2011, \mnras, 413,
  3075, \dodoi{10.1111/j.1365-2966.2011.18381.x}

\bibitem[{{Maguire} {et~al.}(2013){Maguire}, {Sullivan}, {Patat}, {Gal-Yam},
  {Hook}, {Dhawan}, {Howell}, {Mazzali}, {Nugent}, {Pan}, {Podsiadlowski},
  {Simon}, {Sternberg}, {Valenti}, {Baltay}, {Bersier}, {Blagorodnova}, {Chen},
  {Ellman}, {Feindt}, {F{\"o}rster}, {Fraser}, {Gonz{\'a}lez-Gait{\'a}n},
  {Graham}, {Guti{\'e}rrez}, {Hachinger}, {Hadjiyska}, {Inserra}, {Knapic},
  {Laher}, {Leloudas}, {Margheim}, {McKinnon}, {Molinaro}, {Morrell}, {Ofek},
  {Rabinowitz}, {Rest}, {Sand}, {Smareglia}, {Smartt}, {Taddia}, {Walker},
  {Walton}, \& {Young}}]{2013MNRAS.436..222M}
{Maguire}, K., {Sullivan}, M., {Patat}, F., {et~al.} 2013, \mnras, 436, 222,
  \dodoi{10.1093/mnras/stt1586}

\bibitem[{{Mandel} {et~al.}(2014){Mandel}, {Foley}, \&
  {Kirshner}}]{2014ApJ...797...75M}
{Mandel}, K.~S., {Foley}, R.~J., \& {Kirshner}, R.~P. 2014, \apj, 797, 75,
  \dodoi{10.1088/0004-637X/797/2/75}

\bibitem[{{Maoz} {et~al.}(2014){Maoz}, {Mannucci}, \&
  {Nelemans}}]{2014ARA&A..52..107M}
{Maoz}, D., {Mannucci}, F., \& {Nelemans}, G. 2014, \araa, 52, 107,
  \dodoi{10.1146/annurev-astro-082812-141031}

\bibitem[{{Marion} {et~al.}(2016){Marion}, {Brown}, {Vink{\'o}}, {Silverman},
  {Sand}, {Challis}, {Kirshner}, {Wheeler}, {Berlind}, {Brown}, {Calkins},
  {Camacho}, {Dhungana}, {Foley}, {Friedman}, {Graham}, {Howell}, {Hsiao},
  {Irwin}, {Jha}, {Kehoe}, {Macri}, {Maeda}, {Mandel}, {McCully}, {Pandya},
  {Rines}, {Wilhelmy}, \& {Zheng}}]{2016ApJ...820...92M}
{Marion}, G.~H., {Brown}, P.~J., {Vink{\'o}}, J., {et~al.} 2016, \apj, 820, 92,
  \dodoi{10.3847/0004-637X/820/2/92}

\bibitem[{{Maund} {et~al.}(2010){Maund}, {H{\"o}flich}, {Patat}, {Wheeler},
  {Zelaya}, {Baade}, {Wang}, {Clocchiatti}, \& {Quinn}}]{2010ApJ...725L.167M}
{Maund}, J.~R., {H{\"o}flich}, P., {Patat}, F., {et~al.} 2010, \apjl, 725,
  L167, \dodoi{10.1088/2041-8205/725/2/L167}

\bibitem[{{Miller} {et~al.}(2018){Miller}, {Cao}, {Piro}, {Blagorodnova},
  {Bue}, {Cenko}, {Dhawan}, {Ferretti}, {Fox}, {Fremling}, {Goobar}, {Howell},
  {Hosseinzadeh}, {Kasliwal}, {Laher}, {Lunnan}, {Masci}, {McCully}, {Nugent},
  {Sollerman}, {Taddia}, \& {Kulkarni}}]{2018ApJ...852..100M}
{Miller}, A.~A., {Cao}, Y., {Piro}, A.~L., {et~al.} 2018, \apj, 852, 100,
  \dodoi{10.3847/1538-4357/aaa01f}

\bibitem[{{Munari} {et~al.}(2013){Munari}, {Henden}, {Belligoli}, {Castellani},
  {Cherini}, {Righetti}, \& {Vagnozzi}}]{2013NewA...20...30M}
{Munari}, U., {Henden}, A., {Belligoli}, R., {et~al.} 2013, \na, 20, 30,
  \dodoi{10.1016/j.newast.2012.09.003}

\bibitem[{{NASA/IPAC Extragalactic Database
  (NED)}(2019)}]{https://doi.org/10.26132/ned3}
{NASA/IPAC Extragalactic Database (NED)}. 2019, NED Gravitational Wave
  Follow-up (GWF) Service,  IPAC, \dodoi{10.26132/NED3}

\bibitem[{{National Optical Astronomy
  Observatories}(1999)}]{1999ascl.soft11002N}
{National Optical Astronomy Observatories}. 1999, {IRAF: Image Reduction and
  Analysis Facility}, Astrophysics Source Code Library, record ascl:9911.002.
\newblock \doeprint{9911.002}

\bibitem[{{Ni} {et~al.}(2022){Ni}, {Moon}, {Drout}, {Polin}, {Sand},
  {Gonz{\'a}lez-Gait{\'a}n}, {Kim}, {Lee}, {Park}, {Howell}, {Nugent}, {Piro},
  {Brown}, {Galbany}, {Burke}, {Hiramatsu}, {Hosseinzadeh}, {Valenti},
  {Afsariardchi}, {Andrews}, {Antoniadis}, {Arcavi}, {Beaton}, {Bostroem},
  {Carlberg}, {Cenko}, {Cha}, {Dong}, {Gal-Yam}, {Haislip}, {Holoien},
  {Johnson}, {Kouprianov}, {Lee}, {Matzner}, {Morrell}, {McCully}, {Pignata},
  {Reichart}, {Rich}, {Ryder}, {Smith}, {Wyatt}, \&
  {Yang}}]{2022NatAs.tmp...45N}
{Ni}, Y.~Q., {Moon}, D.-S., {Drout}, M.~R., {et~al.} 2022, Nature Astronomy,
  \dodoi{10.1038/s41550-022-01603-4}

\bibitem[{{Nugent} {et~al.}(2011){Nugent}, {Sullivan}, {Cenko}, {Thomas},
  {Kasen}, {Howell}, {Bersier}, {Bloom}, {Kulkarni}, {Kandrashoff},
  {Filippenko}, {Silverman}, {Marcy}, {Howard}, {Isaacson}, {Maguire},
  {Suzuki}, {Tarlton}, {Pan}, {Bildsten}, {Fulton}, {Parrent}, {Sand},
  {Podsiadlowski}, {Bianco}, {Dilday}, {Graham}, {Lyman}, {James}, {Kasliwal},
  {Law}, {Quimby}, {Hook}, {Walker}, {Mazzali}, {Pian}, {Ofek}, {Gal-Yam}, \&
  {Poznanski}}]{2011Natur.480..344N}
{Nugent}, P.~E., {Sullivan}, M., {Cenko}, S.~B., {et~al.} 2011, \nat, 480, 344,
  \dodoi{10.1038/nature10644}

\bibitem[{{Olling} {et~al.}(2015){Olling}, {Mushotzky}, {Shaya}, {Rest},
  {Garnavich}, {Tucker}, {Kasen}, {Margheim}, \&
  {Filippenko}}]{2015Natur.521..332O}
{Olling}, R.~P., {Mushotzky}, R., {Shaya}, E.~J., {et~al.} 2015, \nat, 521,
  332, \dodoi{10.1038/nature14455}

\bibitem[{{Pakmor} {et~al.}(2011){Pakmor}, {Hachinger}, {R{\"o}pke}, \&
  {Hillebrandt}}]{2011A&A...528A.117P}
{Pakmor}, R., {Hachinger}, S., {R{\"o}pke}, F.~K., \& {Hillebrandt}, W. 2011,
  \aap, 528, A117, \dodoi{10.1051/0004-6361/201015653}

\bibitem[{{Pakmor} {et~al.}(2012){Pakmor}, {Kromer}, {Taubenberger}, {Sim},
  {R{\"o}pke}, \& {Hillebrandt}}]{2012ApJ...747L..10P}
{Pakmor}, R., {Kromer}, M., {Taubenberger}, S., {et~al.} 2012, \apjl, 747, L10,
  \dodoi{10.1088/2041-8205/747/1/L10}

\bibitem[{{Pan}(2020)}]{2020ApJ...895L...5P}
{Pan}, Y.-C. 2020, \apjl, 895, L5, \dodoi{10.3847/2041-8213/ab8e47}

\bibitem[{{Pan} {et~al.}(2015){Pan}, {Sullivan}, {Maguire}, {Gal-Yam}, {Hook},
  {Howell}, {Nugent}, \& {Mazzali}}]{2015MNRAS.446..354P}
{Pan}, Y.~C., {Sullivan}, M., {Maguire}, K., {et~al.} 2015, \mnras, 446, 354,
  \dodoi{10.1093/mnras/stu2121}

\bibitem[{{Panessa} \& {Bassani}(2002)}]{2002A&A...394..435P}
{Panessa}, F., \& {Bassani}, L. 2002, \aap, 394, 435,
  \dodoi{10.1051/0004-6361:20021161}

\bibitem[{{Parrent} {et~al.}(2011){Parrent}, {Thomas}, {Fesen}, {Marion},
  {Challis}, {Garnavich}, {Milisavljevic}, {Vink{\`o}}, \&
  {Wheeler}}]{2011ApJ...732...30P}
{Parrent}, J.~T., {Thomas}, R.~C., {Fesen}, R.~A., {et~al.} 2011, \apj, 732,
  30, \dodoi{10.1088/0004-637X/732/1/30}

\bibitem[{{Pastorello} {et~al.}(2007){Pastorello}, {Taubenberger},
  {Elias-Rosa}, {Mazzali}, {Pignata}, {Cappellaro}, {Garavini}, {Nobili},
  {Anupama}, {Bayliss}, {Benetti}, {Bufano}, {Chakradhari}, {Kotak}, {Goobar},
  {Navasardyan}, {Patat}, {Sahu}, {Salvo}, {Schmidt}, {Stanishev}, {Turatto},
  \& {Hillebrandt}}]{2007MNRAS.376.1301P}
{Pastorello}, A., {Taubenberger}, S., {Elias-Rosa}, N., {et~al.} 2007, \mnras,
  376, 1301, \dodoi{10.1111/j.1365-2966.2007.11527.x}

\bibitem[{{Patat} {et~al.}(1997){Patat}, {Barbon}, {Cappellaro}, \&
  {Turatto}}]{1997A&A...317..423P}
{Patat}, F., {Barbon}, R., {Cappellaro}, E., \& {Turatto}, M. 1997, \aap, 317,
  423.
\newblock \doarXiv{astro-ph/9605181}

\bibitem[{{Pereira} {et~al.}(2013){Pereira}, {Thomas}, {Aldering}, {Antilogus},
  {Baltay}, {Benitez-Herrera}, {Bongard}, {Buton}, {Canto}, {Cellier-Holzem},
  {Chen}, {Childress}, {Chotard}, {Copin}, {Fakhouri}, {Fink}, {Fouchez},
  {Gangler}, {Guy}, {Hillebrandt}, {Hsiao}, {Kerschhaggl}, {Kowalski},
  {Kromer}, {Nordin}, {Nugent}, {Paech}, {Pain}, {P{\'e}contal}, {Perlmutter},
  {Rabinowitz}, {Rigault}, {Runge}, {Saunders}, {Smadja}, {Tao},
  {Taubenberger}, {Tilquin}, \& {Wu}}]{2013A&A...554A..27P}
{Pereira}, R., {Thomas}, R.~C., {Aldering}, G., {et~al.} 2013, \aap, 554, A27,
  \dodoi{10.1051/0004-6361/201221008}

\bibitem[{{Perlmutter} {et~al.}(1999){Perlmutter}, {Aldering}, {Goldhaber},
  {Knop}, {Nugent}, {Castro}, {Deustua}, {Fabbro}, {Goobar}, {Groom}, {Hook},
  {Kim}, {Kim}, {Lee}, {Nunes}, {Pain}, {Pennypacker}, {Quimby}, {Lidman},
  {Ellis}, {Irwin}, {McMahon}, {Ruiz-Lapuente}, {Walton}, {Schaefer}, {Boyle},
  {Filippenko}, {Matheson}, {Fruchter}, {Panagia}, {Newberg}, {Couch}, \&
  {Project}}]{1999ApJ...517..565P}
{Perlmutter}, S., {Aldering}, G., {Goldhaber}, G., {et~al.} 1999, \apj, 517,
  565, \dodoi{10.1086/307221}

\bibitem[{{Phillips}(1993)}]{1993ApJ...413L.105P}
{Phillips}, M.~M. 1993, \apjl, 413, L105, \dodoi{10.1086/186970}

\bibitem[{{Phillips} {et~al.}(1999){Phillips}, {Lira}, {Suntzeff}, {Schommer},
  {Hamuy}, \& {Maza}}]{1999AJ....118.1766P}
{Phillips}, M.~M., {Lira}, P., {Suntzeff}, N.~B., {et~al.} 1999, \aj, 118,
  1766, \dodoi{10.1086/301032}

\bibitem[{{Plewa} {et~al.}(2004){Plewa}, {Calder}, \&
  {Lamb}}]{2004ApJ...612L..37P}
{Plewa}, T., {Calder}, A.~C., \& {Lamb}, D.~Q. 2004, \apjl, 612, L37,
  \dodoi{10.1086/424036}

\bibitem[{{Polin} {et~al.}(2019){Polin}, {Nugent}, \&
  {Kasen}}]{2019ApJ...873...84P}
{Polin}, A., {Nugent}, P., \& {Kasen}, D. 2019, \apj, 873, 84,
  \dodoi{10.3847/1538-4357/aafb6a}

\bibitem[{{Prieto} {et~al.}(2006){Prieto}, {Rest}, \&
  {Suntzeff}}]{2006ApJ...647..501P}
{Prieto}, J.~L., {Rest}, A., \& {Suntzeff}, N.~B. 2006, \apj, 647, 501,
  \dodoi{10.1086/504307}

\bibitem[{{Reindl} {et~al.}(2005){Reindl}, {Tammann}, {Sandage}, \&
  {Saha}}]{2005ApJ...624..532R}
{Reindl}, B., {Tammann}, G.~A., {Sandage}, A., \& {Saha}, A. 2005, \apj, 624,
  532, \dodoi{10.1086/429218}

\bibitem[{{Reinecke} {et~al.}(2002){Reinecke}, {Hillebrandt}, \&
  {Niemeyer}}]{2002A&A...391.1167R}
{Reinecke}, M., {Hillebrandt}, W., \& {Niemeyer}, J.~C. 2002, \aap, 391, 1167,
  \dodoi{10.1051/0004-6361:20020885}

\bibitem[{{Richmond} \& {Smith}(2012)}]{2012JAVSO..40..872R}
{Richmond}, M.~W., \& {Smith}, H.~A. 2012, \jaavso, 40, 872.
\newblock \doarXiv{1203.4013}

\bibitem[{{Riess} {et~al.}(1999){Riess}, {Filippenko}, {Li}, \&
  {Schmidt}}]{1999AJ....118.2668R}
{Riess}, A.~G., {Filippenko}, A.~V., {Li}, W., \& {Schmidt}, B.~P. 1999, \aj,
  118, 2668, \dodoi{10.1086/301144}

\bibitem[{{Riess} {et~al.}(1996){Riess}, {Press}, \&
  {Kirshner}}]{1996ApJ...473...88R}
{Riess}, A.~G., {Press}, W.~H., \& {Kirshner}, R.~P. 1996, \apj, 473, 88,
  \dodoi{10.1086/178129}

\bibitem[{{Riess} {et~al.}(1998){Riess}, {Filippenko}, {Challis},
  {Clocchiatti}, {Diercks}, {Garnavich}, {Gilliland}, {Hogan}, {Jha},
  {Kirshner}, {Leibundgut}, {Phillips}, {Reiss}, {Schmidt}, {Schommer},
  {Smith}, {Spyromilio}, {Stubbs}, {Suntzeff}, \&
  {Tonry}}]{1998AJ....116.1009R}
{Riess}, A.~G., {Filippenko}, A.~V., {Challis}, P., {et~al.} 1998, \aj, 116,
  1009, \dodoi{10.1086/300499}

\bibitem[{{Riess} {et~al.}(2016){Riess}, {Macri}, {Hoffmann}, {Scolnic},
  {Casertano}, {Filippenko}, {Tucker}, {Reid}, {Jones}, {Silverman},
  {Chornock}, {Challis}, {Yuan}, {Brown}, \& {Foley}}]{2016ApJ...826...56R}
{Riess}, A.~G., {Macri}, L.~M., {Hoffmann}, S.~L., {et~al.} 2016, \apj, 826,
  56, \dodoi{10.3847/0004-637X/826/1/56}

\bibitem[{{Riess} {et~al.}(2018){Riess}, {Rodney}, {Scolnic}, {Shafer},
  {Strolger}, {Ferguson}, {Postman}, {Graur}, {Maoz}, {Jha}, {Mobasher},
  {Casertano}, {Hayden}, {Molino}, {Hjorth}, {Garnavich}, {Jones}, {Kirshner},
  {Koekemoer}, {Grogin}, {Brammer}, {Hemmati}, {Dickinson}, {Challis}, {Wolff},
  {Clubb}, {Filippenko}, {Nayyeri}, {U}, {Koo}, {Faber}, {Kocevski}, {Bradley},
  \& {Coe}}]{2018ApJ...853..126R}
{Riess}, A.~G., {Rodney}, S.~A., {Scolnic}, D.~M., {et~al.} 2018, \apj, 853,
  126, \dodoi{10.3847/1538-4357/aaa5a9}

\bibitem[{{Robitaille} {et~al.}(2020){Robitaille}, {Deil}, \&
  {Ginsburg}}]{2020ascl.soft11023R}
{Robitaille}, T., {Deil}, C., \& {Ginsburg}, A. 2020, {reproject: Python-based
  astronomical image reprojection}.
\newblock \doeprint{2011.023}

\bibitem[{{R{\"o}pke}(2007)}]{2007ApJ...668.1103R}
{R{\"o}pke}, F.~K. 2007, \apj, 668, 1103, \dodoi{10.1086/520830}

\bibitem[{{Sand} {et~al.}(2018){Sand}, {Graham}, {Boty{\'a}nszki}, {Hiramatsu},
  {McCully}, {Valenti}, {Hosseinzadeh}, {Howell}, {Burke}, {Cartier},
  {Diamond}, {Hsiao}, {Jha}, {Kasen}, {Kumar}, {Marion}, {Suntzeff},
  {Tartaglia}, {Wheeler}, \& {Wyatt}}]{2018ApJ...863...24S}
{Sand}, D.~J., {Graham}, M.~L., {Boty{\'a}nszki}, J., {et~al.} 2018, \apj, 863,
  24, \dodoi{10.3847/1538-4357/aacde8}

\bibitem[{{Schlafly} \& {Finkbeiner}(2011)}]{2011ApJ...737..103S}
{Schlafly}, E.~F., \& {Finkbeiner}, D.~P. 2011, \apj, 737, 103,
  \dodoi{10.1088/0004-637X/737/2/103}

\bibitem[{{Seitenzahl} {et~al.}(2013){Seitenzahl}, {Ciaraldi-Schoolmann},
  {R{\"o}pke}, {Fink}, {Hillebrandt}, {Kromer}, {Pakmor}, {Ruiter}, {Sim}, \&
  {Taubenberger}}]{2013MNRAS.429.1156S}
{Seitenzahl}, I.~R., {Ciaraldi-Schoolmann}, F., {R{\"o}pke}, F.~K., {et~al.}
  2013, \mnras, 429, 1156, \dodoi{10.1093/mnras/sts402}

\bibitem[{{Seitenzahl} {et~al.}(2016){Seitenzahl}, {Kromer}, {Ohlmann},
  {Ciaraldi-Schoolmann}, {Marquardt}, {Fink}, {Hillebrandt}, {Pakmor},
  {R{\"o}pke}, {Ruiter}, {Sim}, \& {Taubenberger}}]{2016A&A...592A..57S}
{Seitenzahl}, I.~R., {Kromer}, M., {Ohlmann}, S.~T., {et~al.} 2016, \aap, 592,
  A57, \dodoi{10.1051/0004-6361/201527251}

\bibitem[{{Shappee} {et~al.}(2018){Shappee}, {Piro}, {Stanek}, {Patel},
  {Margutti}, {Lipunov}, \& {Pogge}}]{2018ApJ...855....6S}
{Shappee}, B.~J., {Piro}, A.~L., {Stanek}, K.~Z., {et~al.} 2018, \apj, 855, 6,
  \dodoi{10.3847/1538-4357/aaa1e9}

\bibitem[{{Shappee} {et~al.}(2019){Shappee}, {Holoien}, {Drout}, {Auchettl},
  {Stritzinger}, {Kochanek}, {Stanek}, {Shaya}, {Narayan}, {ASAS-SN}, {Brown},
  {Bose}, {Bersier}, {Brimacombe}, {Chen}, {Dong}, {Holmbo}, {Katz},
  {Mu{\~n}oz}, {Mutel}, {Post}, {Prieto}, {Shields}, {Tallon}, {Thompson},
  {Vallely}, {Villanueva}, {ATLAS}, {Denneau}, {Flewelling}, {Heinze}, {Smith},
  {Stalder}, {Tonry}, {Weiland}, {Kepler/K2}, {Barclay}, {Barentsen}, {Cody},
  {Dotson}, {Foerster}, {Garnavich}, {Gully-Santiago}, {Hedges}, {Howell},
  {Kasen}, {Margheim}, {Mushotzky}, {Rest}, {Tucker}, {Villar}, {Zenteno},
  {Kepler Spacecraft Team}, {Beerman}, {Bjella}, {Castillo}, {Coughlin},
  {Elsaesser}, {Flynn}, {Gangopadhyay}, {Griest}, {Hanley}, {Kampmeier},
  {Kloetzel}, {Kohnert}, {Labonde}, {Larsen}, {Larson}, {McCalmont-Everton},
  {McGinn}, {Migliorini}, {Moffatt}, {Muszynski}, {Nystrom}, {Osborne},
  {Packard}, {Peterson}, {Redick}, {Reedy}, {Ross}, {Spencer}, {Steward}, {Van
  Cleve}, {Cardoso}, {Weschler}, {Wheaton}, {Pan-STARRS}, {Bulger}, {Chambers},
  {Flewelling}, {Huber}, {Lowe}, {Magnier}, {Schultz}, {Waters}, {Willman},
  {PTSS/TNTS}, {Baron}, {Chen}, {Derkacy}, {Huang}, {Li}, {Li}, {Li}, {Mo},
  {Rui}, {Sai}, {Wang}, {Wang}, {Wang}, {Xiang}, {Zhang}, {Zhang}, {Zhang},
  {Zhang}, {Zhang}, {Zhao}, {Brown}, {Hermes}, {Nordin}, {Points}, {S{\'o}dor},
  {Strampelli}, \& {Zenteno}}]{2019ApJ...870...13S}
{Shappee}, B.~J., {Holoien}, T.~W.~S., {Drout}, M.~R., {et~al.} 2019, \apj,
  870, 13, \dodoi{10.3847/1538-4357/aaec79}

\bibitem[{{Shen} {et~al.}(2021){Shen}, {Blondin}, {Kasen}, {Dessart},
  {Townsley}, {Boos}, \& {Hillier}}]{2021ApJ...909L..18S}
{Shen}, K.~J., {Blondin}, S., {Kasen}, D., {et~al.} 2021, \apjl, 909, L18,
  \dodoi{10.3847/2041-8213/abe69b}

\bibitem[{{Silverman} {et~al.}(2012){Silverman}, {Foley}, {Filippenko},
  {Ganeshalingam}, {Barth}, {Chornock}, {Griffith}, {Kong}, {Lee}, {Leonard},
  {Matheson}, {Miller}, {Steele}, {Barris}, {Bloom}, {Cobb}, {Coil},
  {Desroches}, {Gates}, {Ho}, {Jha}, {Kandrashoff}, {Li}, {Mandel}, {Modjaz},
  {Moore}, {Mostardi}, {Papenkova}, {Park}, {Perley}, {Poznanski}, {Reuter},
  {Scala}, {Serduke}, {Shields}, {Swift}, {Tonry}, {Van Dyk}, {Wang}, \&
  {Wong}}]{2012MNRAS.425.1789S}
{Silverman}, J.~M., {Foley}, R.~J., {Filippenko}, A.~V., {et~al.} 2012, \mnras,
  425, 1789, \dodoi{10.1111/j.1365-2966.2012.21270.x}

\bibitem[{{Silverman} {et~al.}(2013){Silverman}, {Nugent}, {Gal-Yam},
  {Sullivan}, {Howell}, {Filippenko}, {Arcavi}, {Ben-Ami}, {Bloom}, {Cenko},
  {Cao}, {Chornock}, {Clubb}, {Coil}, {Foley}, {Graham}, {Griffith}, {Horesh},
  {Kasliwal}, {Kulkarni}, {Leonard}, {Li}, {Matheson}, {Miller}, {Modjaz},
  {Ofek}, {Pan}, {Perley}, {Poznanski}, {Quimby}, {Steele}, {Sternberg}, {Xu},
  \& {Yaron}}]{2013ApJS..207....3S}
{Silverman}, J.~M., {Nugent}, P.~E., {Gal-Yam}, A., {et~al.} 2013, \apjs, 207,
  3, \dodoi{10.1088/0067-0049/207/1/3}

\bibitem[{{Stahl} {et~al.}(2019){Stahl}, {Zheng}, {de Jaeger}, {Filippenko},
  {Bigley}, {Blanchard}, {Blanchard}, {Brink}, {Cargill}, {Casper}, {Channa},
  {Choi}, {Choksi}, {Chu}, {Clubb}, {Cohen}, {Ellison}, {Falcon}, {Fazeli},
  {Fuller}, {Ganeshalingam}, {Gates}, {Gould}, {Halevi}, {Hayakawa},
  {Hestenes}, {Jeffers}, {Joubert}, {Kandrashoff}, {Kim}, {Kim}, {Kislak},
  {Kleiser}, {Kong}, {de Kouchkovsky}, {Krishnan}, {Kumar}, {Leja}, {Leonard},
  {Li}, {Li}, {Lu}, {Mason}, {Molloy}, {Pina}, {Rex}, {Ross}, {Stegman},
  {Tang}, {Thrasher}, {Wang}, {Wilkins}, {Yuk}, {Yunus}, \&
  {Zhang}}]{2019MNRAS.490.3882S}
{Stahl}, B.~E., {Zheng}, W., {de Jaeger}, T., {et~al.} 2019, \mnras, 490, 3882,
  \dodoi{10.1093/mnras/stz2742}

\bibitem[{{Stanishev} {et~al.}(2007){Stanishev}, {Goobar}, {Benetti}, {Kotak},
  {Pignata}, {Navasardyan}, {Mazzali}, {Amanullah}, {Garavini}, {Nobili},
  {Qiu}, {Elias-Rosa}, {Ruiz-Lapuente}, {Mendez}, {Meikle}, {Patat},
  {Pastorello}, {Altavilla}, {Gustafsson}, {Harutyunyan}, {Iijima},
  {Jakobsson}, {Kichizhieva}, {Lundqvist}, {Mattila}, {Melinder}, {Pavlenko},
  {Pavlyuk}, {Sollerman}, {Tsvetkov}, {Turatto}, \&
  {Hillebrandt}}]{2007A&A...469..645S}
{Stanishev}, V., {Goobar}, A., {Benetti}, S., {et~al.} 2007, \aap, 469, 645,
  \dodoi{10.1051/0004-6361:20066020}

\bibitem[{{Sternberg} {et~al.}(2011){Sternberg}, {Gal-Yam}, {Simon}, {Leonard},
  {Quimby}, {Phillips}, {Morrell}, {Thompson}, {Ivans}, {Marshall},
  {Filippenko}, {Marcy}, {Bloom}, {Patat}, {Foley}, {Yong}, {Penprase},
  {Beeler}, {Allende Prieto}, \& {Stringfellow}}]{2011Sci...333..856S}
{Sternberg}, A., {Gal-Yam}, A., {Simon}, J.~D., {et~al.} 2011, Science, 333,
  856, \dodoi{10.1126/science.1203836}

\bibitem[{{Takanashi} {et~al.}(2008){Takanashi}, {Doi}, \&
  {Yasuda}}]{2008MNRAS.389.1577T}
{Takanashi}, N., {Doi}, M., \& {Yasuda}, N. 2008, \mnras, 389, 1577,
  \dodoi{10.1111/j.1365-2966.2008.13694.x}

\bibitem[{{Thomas} {et~al.}(2011){Thomas}, {Aldering}, {Antilogus}, {Aragon},
  {Bailey}, {Baltay}, {Bongard}, {Buton}, {Canto}, {Childress}, {Chotard},
  {Copin}, {Fakhouri}, {Gangler}, {Hsiao}, {Kerschhaggl}, {Kowalski}, {Loken},
  {Nugent}, {Paech}, {Pain}, {Pecontal}, {Pereira}, {Perlmutter}, {Rabinowitz},
  {Rigault}, {Rubin}, {Runge}, {Scalzo}, {Smadja}, {Tao}, {Weaver}, {Wu},
  {Brown}, {Milne}, \& {Nearby Supernova Factory}}]{2011ApJ...743...27T}
{Thomas}, R.~C., {Aldering}, G., {Antilogus}, P., {et~al.} 2011, \apj, 743, 27,
  \dodoi{10.1088/0004-637X/743/1/27}

\bibitem[{{Th{\"o}ne} {et~al.}(2009){Th{\"o}ne}, {Micha{\l}owski}, {Leloudas},
  {Cox}, {Fynbo}, {Sollerman}, {Hjorth}, \& {Vreeswijk}}]{2009ApJ...698.1307T}
{Th{\"o}ne}, C.~C., {Micha{\l}owski}, M.~J., {Leloudas}, G., {et~al.} 2009,
  \apj, 698, 1307, \dodoi{10.1088/0004-637X/698/2/1307}

\bibitem[{{Tomasella} {et~al.}(2021){Tomasella}, {Benetti}, {Cappellaro}, \&
  {Pastorello}}]{2021TNSCR1031....1T}
{Tomasella}, L., {Benetti}, S., {Cappellaro}, E., \& {Pastorello}, A. 2021,
  Transient Name Server Classification Report, 2021-1031, 1

\bibitem[{{Tsvetkov} \& {Elenin}(2010)}]{2010PZ.....30....2T}
{Tsvetkov}, D.~Y., \& {Elenin}, L. 2010, Peremennye Zvezdy, 30, 2.
\newblock \doarXiv{1003.2558}

\bibitem[{{Tsvetkov} {et~al.}(2013){Tsvetkov}, {Shugarov}, {Volkov},
  {Goranskij}, {Pavlyuk}, {Katysheva}, {Barsukova}, \&
  {Valeev}}]{2013CoSka..43...94T}
{Tsvetkov}, D.~Y., {Shugarov}, S.~Y., {Volkov}, I.~M., {et~al.} 2013,
  Contributions of the Astronomical Observatory Skalnate Pleso, 43, 94.
\newblock \doarXiv{1311.3484}

\bibitem[{{Tully} {et~al.}(2013){Tully}, {Courtois}, {Dolphin}, {Fisher},
  {H{\'e}raudeau}, {Jacobs}, {Karachentsev}, {Makarov}, {Makarova},
  {Mitronova}, {Rizzi}, {Shaya}, {Sorce}, \& {Wu}}]{2013AJ....146...86T}
{Tully}, R.~B., {Courtois}, H.~M., {Dolphin}, A.~E., {et~al.} 2013, \aj, 146,
  86, \dodoi{10.1088/0004-6256/146/4/86}

\bibitem[{{van Dokkum}(2001)}]{2001PASP..113.1420V}
{van Dokkum}, P.~G. 2001, \pasp, 113, 1420, \dodoi{10.1086/323894}

\bibitem[{Virtanen {et~al.}(2020)Virtanen, Gommers, Oliphant, Haberland, Reddy,
  Cournapeau, Burovski, Peterson, Weckesser, Bright, {van der Walt}, Brett,
  Wilson, Millman, Mayorov, Nelson, Jones, Kern, Larson, Carey, Polat, Feng,
  Moore, {VanderPlas}, Laxalde, Perktold, Cimrman, Henriksen, Quintero, Harris,
  Archibald, Ribeiro, Pedregosa, {van Mulbregt}, \& {SciPy 1.0
  Contributors}}]{2020SciPy-NMeth}
Virtanen, P., Gommers, R., Oliphant, T.~E., {et~al.} 2020, Nature Methods, 17,
  261, \dodoi{10.1038/s41592-019-0686-2}

\bibitem[{{Wang} {et~al.}(2020){Wang}, {Contreras}, {Hu}, {Hamuy}, {Hsiao},
  {Sand}, {Anderson}, {Ashall}, {Burns}, {Chen}, {Diamond}, {Davis},
  {F{\"o}rster}, {Galbany}, {Gonz{\'a}lez-Gait{\'a}n}, {Gromadzki}, {Hoeflich},
  {Li}, {Marion}, {Morrell}, {Pignata}, {Prieto}, {Phillips}, {Shahbandeh},
  {Suntzeff}, {Valenti}, {Wang}, {Wang}, {Young}, {Yu}, \&
  {Zhang}}]{2020ApJ...904...14W}
{Wang}, L., {Contreras}, C., {Hu}, M., {et~al.} 2020, \apj, 904, 14,
  \dodoi{10.3847/1538-4357/abba82}

\bibitem[{{Wang} {et~al.}(2013){Wang}, {Wang}, {Filippenko}, {Zhang}, \&
  {Zhao}}]{2013Sci...340..170W}
{Wang}, X., {Wang}, L., {Filippenko}, A.~V., {Zhang}, T., \& {Zhao}, X. 2013,
  Science, 340, 170, \dodoi{10.1126/science.1231502}

\bibitem[{{Wang} {et~al.}(2006){Wang}, {Wang}, {Pain}, {Zhou}, \&
  {Li}}]{2006ApJ...645..488W}
{Wang}, X., {Wang}, L., {Pain}, R., {Zhou}, X., \& {Li}, Z. 2006, \apj, 645,
  488, \dodoi{10.1086/504312}

\bibitem[{{Wang} {et~al.}(2005){Wang}, {Wang}, {Zhou}, {Lou}, \&
  {Li}}]{2005ApJ...620L..87W}
{Wang}, X., {Wang}, L., {Zhou}, X., {Lou}, Y.-Q., \& {Li}, Z. 2005, \apjl, 620,
  L87, \dodoi{10.1086/428774}

\bibitem[{{Wang} {et~al.}(2009){Wang}, {Filippenko}, {Ganeshalingam}, {Li},
  {Silverman}, {Wang}, {Chornock}, {Foley}, {Gates}, {Macomber}, {Serduke},
  {Steele}, \& {Wong}}]{2009ApJ...699L.139W}
{Wang}, X., {Filippenko}, A.~V., {Ganeshalingam}, M., {et~al.} 2009, \apjl,
  699, L139, \dodoi{10.1088/0004-637X/699/2/L139}

\bibitem[{{Webbink}(1984)}]{1984ApJ...277..355W}
{Webbink}, R.~F. 1984, \apj, 277, 355, \dodoi{10.1086/161701}

\bibitem[{{Whelan} \& {Iben}(1973)}]{1973ApJ...186.1007W}
{Whelan}, J., \& {Iben}, Icko, J. 1973, \apj, 186, 1007, \dodoi{10.1086/152565}

\bibitem[{{Wong} {et~al.}(2020){Wong}, {Suyu}, {Chen}, {Rusu}, {Millon},
  {Sluse}, {Bonvin}, {Fassnacht}, {Taubenberger}, {Auger}, {Birrer}, {Chan},
  {Courbin}, {Hilbert}, {Tihhonova}, {Treu}, {Agnello}, {Ding}, {Jee},
  {Komatsu}, {Shajib}, {Sonnenfeld}, {Blandford}, {Koopmans}, {Marshall}, \&
  {Meylan}}]{2020MNRAS.498.1420W}
{Wong}, K.~C., {Suyu}, S.~H., {Chen}, G. C.~F., {et~al.} 2020, \mnras, 498,
  1420, \dodoi{10.1093/mnras/stz3094}

\bibitem[{{Wu} {et~al.}(1995){Wu}, {Yan}, \& {Zou}}]{1995A&A...294L...9W}
{Wu}, H., {Yan}, H.~J., \& {Zou}, Z.~L. 1995, \aap, 294, L9

\bibitem[{{Yang} {et~al.}(2020){Yang}, {Hoeflich}, {Baade}, {Maund}, {Wang},
  {Brown}, {Stevance}, {Arcavi}, {Burke}, {Cikota}, {Clocchiatti}, {Gal-Yam},
  {Graham}, {Hiramatsu}, {Hosseinzadeh}, {Howell}, {Jha}, {McCully}, {Patat},
  {Sand}, {Schulze}, {Spyromilio}, {Valenti}, {Vink{\'o}}, {Wang}, {Wheeler},
  {Yaron}, \& {Zhang}}]{2020ApJ...902...46Y}
{Yang}, Y., {Hoeflich}, P., {Baade}, D., {et~al.} 2020, \apj, 902, 46,
  \dodoi{10.3847/1538-4357/aba759}

\bibitem[{{Yaron} \& {Gal-Yam}(2012)}]{2012PASP..124..668Y}
{Yaron}, O., \& {Gal-Yam}, A. 2012, \pasp, 124, 668, \dodoi{10.1086/666656}

\bibitem[{{Zeng} {et~al.}(2021){Zeng}, {Wang}, {Esamdin}, {Pellegrino},
  {Burke}, {Stahl}, {Zheng}, {Filippenko}, {Howell}, {Sand}, {Valenti}, {Mo},
  {Xi}, {Liu}, {Zhang}, {Li}, {Iskandar}, {Zhang}, {Lin}, {Sai}, {Xiang},
  {Wei}, {Zhang}, {Reichart}, {Brink}, {McCully}, {Hiramatsu}, {Hosseinzadeh},
  {Jeffers}, {Ross}, {Stegman}, {Wang}, {Zhang}, \& {Ma}}]{2021ApJ...919...49Z}
{Zeng}, X., {Wang}, X., {Esamdin}, A., {et~al.} 2021, \apj, 919, 49,
  \dodoi{10.3847/1538-4357/ac0e9c}

\bibitem[{{Zhang} {et~al.}(2017){Zhang}, {Childress}, {Davis}, {Karpenka},
  {Lidman}, {Schmidt}, \& {Smith}}]{2017MNRAS.471.2254Z}
{Zhang}, B.~R., {Childress}, M.~J., {Davis}, T.~M., {et~al.} 2017, \mnras, 471,
  2254, \dodoi{10.1093/mnras/stx1600}

\end{thebibliography}
\bibliographystyle{aasjournal}

\begin{deluxetable}{ccccccccccccccc}
\tablecaption{$BVRI$ photometry of SN 2021hpr \label{tab:phot}}
\tablehead{\colhead{UT Date} & \colhead{MJD} & \colhead{phase\tablenotemark{a}} & \colhead{$t_{\text{Bexp}}$} & \colhead{$B$} & \colhead{$\delta B$} & \colhead{$t_{\text{Vexp}}$} & \colhead{$V$} & \colhead{$\delta V$} & \colhead{$t_{\text{Rexp}}$} & \colhead{$R$} & \colhead{$\delta R$} & \colhead{$t_{\text{Iexp}}$} & \colhead{$I$} & \colhead{$\delta I$}\\ 
\colhead{ } & \colhead{ } & \colhead{$\mathrm{days}$} & \colhead{$\mathrm{s}$} & \colhead{$\mathrm{mag}$} & \colhead{$\mathrm{mag}$} & \colhead{$\mathrm{s}$} & \colhead{$\mathrm{mag}$} & \colhead{$\mathrm{mag}$} & \colhead{$\mathrm{s}$} & \colhead{$\mathrm{mag}$} & \colhead{$\mathrm{mag}$} & \colhead{$\mathrm{s}$} & \colhead{$\mathrm{mag}$} & \colhead{$\mathrm{mag}$}}
\startdata
2021 April 03 & 59,307.521 & $-$14.368 & 180.0 & 17.791 & 0.056 & 180.0 & 17.252 & 0.041 & 180.0 & 17.206 & 0.038 & 180.0 & 17.045 & 0.062 \\
2021 April 04 & 59,308.485 & $-$13.404 & 180.0 & 17.253 & 0.105 & 180.0 & 16.597 & 0.057 & 180.0 & 16.580 & 0.045 & 180.0 & 16.909 & 0.099 \\
2021 April 07 & 59,311.569 & $-$10.320 & 200.0 & 15.397 & 0.016 & 200.0 & 15.294 & 0.015 & 200.0 & 15.321 & 0.013 & 200.0 & 15.338 & 0.017 \\
2021 April 08 & 59,312.505 & $-$9.384 & 200.0 & 15.127 & 0.021 & 200.0 & 15.045 & 0.026 & 200.0 & 15.108 & 0.017 & 200.0 & 15.122 & 0.018 \\
2021 April 09 & 59,313.559 & $-$8.330 & 200.0 & 14.832 & 0.059 & 200.0 & 14.870 & 0.040 & 200.0 & 14.871 & 0.024 & 200.0 & 14.859 & 0.022 \\
2021 April 10 & 59,314.566 & $-$7.323 & 200.0 & 14.565 & 0.075 & 200.0 & 14.601 & 0.065 & 200.0 & 14.670 & 0.038 & 200.0 & 14.692 & 0.039 \\
2021 April 13 & 59,317.508 & $-$4.381 & 200.0 & 14.352 & 0.012 & 200.0 & 14.365 & 0.014 & 200.0 & 14.379 & 0.014 & 200.0 & 14.500 & 0.015 \\
2021 April 14 & 59,318.498 & $-$3.391 & 200.0 & 14.297 & 0.028 & 200.0 & 14.278 & 0.020 & 200.0 & 14.334 & 0.013 & 200.0 & 14.479 & 0.016 \\
2021 April 18 & 59,322.506 & 0.617 & 200.0 & 14.202 & 0.030 & 200.0 & 14.167 & 0.022 & 200.0 & 14.185 & 0.015 & 200.0 & 14.487 & 0.018 \\
2021 April 19 & 59,323.506 & 1.617 & 200.0 & 14.200 & 0.034 & 200.0 & 14.162 & 0.023 & 200.0 & 14.172 & 0.017 & 200.0 & 14.524 & 0.020 \\
2021 April 20 & 59,324.493 & 2.604 & 200.0 & 14.263 & 0.117 & 200.0 & 14.141 & 0.057 & 200.0 & 14.180 & 0.043 & 200.0 & 14.558 & 0.035 \\
2021 April 27 & 59,331.727 & 9.838 & -- & -- & -- & 200.0 & 14.332 & 0.018 & 200.0 & 14.480 & 0.015 & 200.0 & 14.818 & 0.017 \\
2021 April 30 & 59,334.509 & 12.620 & 200.0 & 14.925 & 0.017 & 200.0 & 14.560 & 0.021 & 200.0 & 14.700 & 0.013 & 200.0 & 14.950 & 0.017 \\
2021 May 2 & 59,336.507 & 14.618 & 200.0 & 15.085 & 0.054 & 200.0 & 14.612 & 0.058 & 200.0 & 14.794 & 0.030 & 200.0 & 15.044 & 0.025 \\
2021 May 5 & 59,339.511 & 17.622 & 200.0 & 15.412 & 0.039 & 200.0 & 14.873 & 0.027 & 200.0 & 14.920 & 0.016 & 200.0 & 14.939 & 0.020 \\
2021 May 7 & 59,341.519 & 19.630 & 200.0 & 15.775 & 0.046 & 200.0 & 14.980 & 0.032 & 200.0 & 14.932 & 0.023 & 200.0 & 14.847 & 0.030 \\
2021 May 8 & 59,342.521 & 20.632 & 200.0 & 15.831 & 0.023 & 200.0 & 15.016 & 0.014 & 200.0 & 14.934 & 0.011 & 200.0 & 14.840 & 0.013 \\
2021 May 10 & 59,344.559 & 22.670 & 200.0 & 16.017 & 0.019 & 200.0 & 15.131 & 0.018 & 200.0 & 14.963 & 0.012 & 200.0 & 14.775 & 0.017 \\
2021 May 11 & 59,345.528 & 23.639 & 200.0 & 16.132 & 0.043 & 200.0 & 15.196 & 0.024 & 200.0 & 14.991 & 0.017 & 200.0 & 14.760 & 0.026 \\
2021 May 16 & 59,350.526 & 28.637 & 200.0 & 16.493 & 0.048 & 200.0 & 15.488 & 0.026 & 200.0 & 15.124 & 0.018 & 200.0 & 14.758 & 0.024 \\
2021 May 17 & 59,351.522 & 29.633 & 200.0 & 16.699 & 0.084 & 200.0 & 15.543 & 0.030 & 200.0 & 15.195 & 0.025 & 200.0 & 14.747 & 0.020 \\
2021 May 24 & 59,358.548 & 36.659 & 200.0 & 17.063 & 0.055 & 200.0 & 16.013 & 0.024 & 200.0 & 15.658 & 0.021 & 200.0 & 15.230 & 0.023 \\
2021 May 25 & 59,359.550 & 37.661 & -- & -- & -- & 200.0 & 16.030 & 0.042 & 200.0 & 15.690 & 0.032 & 200.0 & 15.289 & 0.032 \\
2021 May 30 & 59,364.597 & 42.708 & 200.0 & 17.361 & 0.034 & 200.0 & 16.267 & 0.014 & 200.0 & 15.954 & 0.013 & 200.0 & 15.605 & 0.019 \\
2021 June 2 & 59,367.589 & 45.700 & 200.0 & 17.399 & 0.029 & 200.0 & 16.332 & 0.015 & 200.0 & 16.021 & 0.013 & 200.0 & 15.727 & 0.017 \\
2021 June 3 & 59,368.557 & 46.668 & -- & -- & -- & 200.0 & 16.397 & 0.018 & 200.0 & 16.060 & 0.015 & 200.0 & 15.752 & 0.021
\enddata
\tablenotetext{a}{Days relative to the $B$-band maximum (MJD = 59,321.862 $\pm$ 0.450)}
\end{deluxetable}

\begin{deluxetable}{ccccccc}
\tablecaption{Log of spectroscopic observations of SN 2021hpr \label{tab:spec}}
\tablehead{\colhead{UT Date} & \colhead{MJD} & \colhead{Phase\tablenotemark{a}} & \colhead{Range} & \colhead{Resolution} & \colhead{$t_{exp}$} & \colhead{Airmass}\\
\colhead{ } & \colhead{ } & \colhead{days} & \colhead{$\mathrm{\AA}$} & \colhead{$\mathrm{\AA}$} & \colhead{$\mathrm{s}$} & \colhead{ }}
\startdata
April 03, 2021 & 59307.517 & $-$14.372 & 3700-8800 & 20 & 3 $\times$ 1100.0 & 1.200 \\
April 13, 2021 & 59317.584 & $-$4.305 & 3700-8800 & 20 & 3 $\times$ 1200.0 & 1.209 \\
May 01, 2021 & 59335.501 & 13.612 & 3700-8800 & 20 & 3 $\times$ 1200.0 & 1.193 \\
May 10, 2021 & 59344.536 & 22.647 & 3700-8800 & 20 & 3 $\times$ 1200.0 & 1.231 \\
June 03, 2021 & 59368.581 & 46.692 & 3700-8800 & 20 & 3 $\times$ 1200.0 & 1.427 \\
June 20, 2021 & 59385.565 & 63.676 & 3700-8800 & 20 & 3 $\times$ 1200.0 & 1.515
\enddata
\tablenotetext{a}{Days relative to the $B$-band maximum (MJD = 59321.862 $\pm$ 0.450)}
\end{deluxetable}

\begin{deluxetable}{ccc}
\tablecaption{Peak magnitudes and corresponding MJDs of SN 2021hpr in $BVRI$ bands\label{tab:mjd m in peak}}
\tablehead{\colhead{Band} & \colhead{$\text{MJD}_{max}$} & \colhead{$m_{max}$}\\
\colhead{ } &
\colhead{ } & 
\colhead{mag}}
\startdata
$B$ & 59,321.862 $\pm$ 0.450 & 14.017 $\pm$ 0.017 \\
$V$ & 59,323.220 $\pm$ 0.733 & 14.091 $\pm$ 0.014 \\
$R$ & 59,323.522 $\pm$ 0.537 & 14.123 $\pm$ 0.009 \\
$I$ & 59,320.327 $\pm$ 0.504 & 14.415 $\pm$ 0.010
\enddata
\end{deluxetable}

\begin{deluxetable}{cccc}
\tablecaption{Supernovae observed in NGC 3147 
\label{tab:SNe in NGC 3147}}
\tablehead{\colhead{Name} & \colhead{Type} & \colhead{R.A.(J2000)} & \colhead{Decl.(J2000)}}
\startdata
SN 1972H & SN Ia & 10:17:00.400 & +73:24:39.00 \\
SN 1997bq & SN Ia & 10:17:05.330 & +73:23:02.11 \\
SN 2006gi & SN Ib & 10:16:46.759 & +73:26:26.41 \\
SN 2008fv & SN Ia & 10:16:57.281 & +73:24:36.40 \\
SN 2021do & SN Ic & 10:16:56.522 & +73:23:50.93 \\
SN 2021hpr & SN Ia & 10:16:38.680 & +73:24:01.80
\enddata
\end{deluxetable}

\begin{deluxetable}{ccc}
\tablecaption{Photometric Parameters of SN 2021hpr \label{tab:param}}
\tablehead{\colhead{Parameter} & \colhead{Value} & \colhead{Reference}}
\startdata
RA & 10:16:38.680 & 1 \\
Dec & +73:24:01.80 & 1 \\
Type & SN Ia & 1,3 \\
Host galaxy & NGC 3147 & 1 \\
Discoverer & Koichi Itagaki & 1 \\
Discovery Date & 2021 04 02 10:46:28.000 & 1 \\
Discovery Mag\tablenotemark{a} & 17.7 mag & 1 \\
redshift & 0.009346 & 1 \\
Distance & 46.02 $\pm$ 3.23 Mpc & 3 \\
Distance modulus & 33.46 $\pm$ 0.21 mag & 3 \\
$E(B-V)_{gal}$ & 0.0208 mag & 2 \\
$A_B$ & 0.088 mag & 2 \\
$A_V$ & 0.067 mag & 2 \\
$A_R$ & 0.053 mag & 2 \\
$A_I$ & 0.037 mag & 2 \\
$t_{max}^B$ & $59321.862\pm 0.450$ & 3 \\
$m_{max}^B$ & $14.017\pm 0.017$ mag & 3 \\
$M_{max}^B$ & $-19.531\pm 0.210$ mag & 3 \\
$\Delta m_{15}(B)$\tablenotemark{b} & $0.949\pm 0.019$ mag & 3 \\
$t_0^B$ & $59305.438\pm 0.450$ & 3 \\
$t_{rise}^B$ & $16.424\pm 0.078$ d & 3
\enddata
\tablenotetext{a}{Unfiltered}
\tablenotetext{b}{Derived by SNooPy \citep{2011AJ....141...19B, 2014ApJ...789...32B}}
\tablerefs{1 - \cite{2021TNSTR.998....1I}; 2 - \cite{2011ApJ...737..103S}; 3 - this paper}
\end{deluxetable}

\end{document}